\documentclass[twocolumn,english,aps,prb,showpacs,floatfix,,superscriptaddress]{revtex4-1}
\usepackage[latin9]{inputenc}
\usepackage{amsmath}
\usepackage{amssymb}
\usepackage{wasysym}
\usepackage{graphicx}
\usepackage{esint}
\usepackage{babel}

\begin{document}

\title{Conditional counting statistics of electrons tunneling through quantum
dot systems measured by a quantum point contact}

\author{Yen-Jui Chang}
\affiliation{Department of Physics and Center for Theoretical Sciences, National Taiwan University, Taipei 10617, Taiwan}
\affiliation{Center for Quantum Science and Engineering, National
  Taiwan University, Taipei 10617, Taiwan}
\author{Tsung-Kang Yeh}
\affiliation{Department of Physics and Center for Theoretical
  Sciences, National Taiwan University, Taipei 10617, Taiwan}
\author{Chao-Hung Wan}
\affiliation{Department of Physics and Center for Theoretical Sciences, National Taiwan University, Taipei 10617, Taiwan}
\author{D. Wahyu Utami}
\affiliation{Research Systems and Reporting, Australian Catholic
  University, North Sydney, NSW 2060, Australia}
\author{Gerard J. Milburn}
\affiliation{Centre for Engineered Quantum Systems, School of
  Mathematics and Physics, The University of Queensland, St Lucia, QLD 4072, Australia}
\author{Hsi-Sheng Goan}
\email{goan@phys.ntu.edu.tw}
\affiliation{Department of Physics and Center for Theoretical Sciences, National Taiwan University, Taipei 10617, Taiwan}
\affiliation{Center for Quantum Science and Engineering, National Taiwan University, Taipei 10617, Taiwan}
\date{\today}

\begin{abstract}
We theoretically study the conditional counting statistics of
electron transport through 
a system consisting of a single quantum dot (SQD) or coherently coupled
double quantum dots 
(DQD's) monitored by a nearby quantum point contact (QPC)
using 
the generating functional approach with 
the maximum eigenvalue
of the evolution equation matrix method, the quantum trajectory
theory method (Monte Carlo method),
and an efficient method we develop.
The conditional current cumulants
that are significantly different from their unconditional counterparts can provide additional information and insight into the electron 
transport properties of mesoscopic nanostructure systems.  
The efficient method we develop for calculating the conditional
counting statistics 
is numerically stable,
and is capable of calculating
the conditional counting statistics for a more complex system than the maximum eigenvalue method
and for a wider range of parameters than the quantum trajectory method. 
We apply our method to investigate how the QPC shot noise affects the
conditional counting statistics of the SQD system, going beyond the
treatment and 
parameter regime studied in the literature.
We also investigate the case when the
interdot coherent coupling is comparable to the dephasing rate caused
by the back action of the QPC in the DQD system, in which 
there is considerable discrepancy in the calculated conditional 
current cumulants 
between the population rate (master-) equation approach of sequential
tunneling and the full quantum master-equation approach of coherent
tunneling.

\end{abstract}


\maketitle

\section{introduction}
The time-resolved measurement of electron charges through a single quantum dot
(SQD) by a nearby  quantum point contact (QPC) detector
has been demonstrated experimentally \cite{Elzerman04,Schleser04,Vandersypen04,Gustavsson06,Gustavsson09,Ubbelohde12}.
The ability to detect individual
charges in real time makes it possible to count electrons one by one
as they pass through the quantum dot(QD) \cite{Elzerman04,Schleser04,Vandersypen04,Gustavsson06,Gustavsson09,Ubbelohde12,Lu03,Fujisawa04,Bylander05,Fujisawa06,Fricke07}. 
The time-resolved charge detection has allowed the precise measurement of the QD
shot noise at subfemtoampere current levels, 
and the full counting statistics (FCS) of the current \cite{Gustavsson06,Gustavsson09,Ubbelohde12}.  

FCS in quantum transport provides information of quantum statistical properties of transport
phenomena and is studied mostly based on the computation
of its moment or cumulant generating function \cite{Levitov96,key-4,Nazarov03}.
Computing the generating function is more convenient in practice 
than the direct calculation of the probability distribution function and then performing average over the
powers of electron number or current. 
A theoretical approach called number-resolved master-equation
approach has been applied to calculate the generating functions and 
unconditional 
FCS for the nanostructure
electron transport problems
\cite{key-4,Nazarov03,Gurvitz97,Schoell04,key-7,key-8,key-6}.

When a measurement is made on a single quantum system
and the result is available, the state or density matrix of the system
is a conditional state conditioned on the measurement result 
\cite{Korotkov01,Goan01a,Goan01b}. Thus,
the conditional state of the system is important
when its subsequent time evolution is concerned. If a single system
is under continuous monitoring and one wants to map out the system
state evolution conditioned on the continuous in time measurement
results, the conditional (Bayesian) stochastic Schr\"dinger or
stochastic master 
equation approach or the quantum trajectory theory (quantum Monte Carlo method)
can be employed
\cite{Korotkov01,Goan01a,Goan01b,Goan03,Goan04}.
Each quantum trajectory 
can mimic the stochastic system state evolution 
conditioned on the  continuous in time measurement outcomes in
a single run of a realistic experiment. 
The stochastic element
in the quantum trajectory corresponds exactly to the consequence of
the random outcomes of 
the measurement record \cite{Korotkov01,Goan01a,Goan01b,Goan03,Goan04}.
Thus, the quantum trajectories 
have the full information of the statistical properties about the
measured system and can give insight to the unconditional quantities.

In some cases, one is concerned with the system state or physical observables 
conditioned on some average quantities (e.g., average current) in a
given period of time 
rather than instantaneous and continuous in time measurement results.  
For example, the conditional counting statistics of electron
transport through a 
SQD coupling to a QPC
has been measured in the experiment by Sukhorukov {\it et
al}. \cite{Sukhorukov07}. 
The conditional FCS that is the statistical current cumulants of
one system given the observation of a particular average current in
time $t$ in the other
system could be substantially different from its unconditional
counterpart. 
A theoretical approach that utilizes the number-resolved rate (master)
equation of a bistable SQD system and neglects the QPC shot
noise 
was put forward to calculate the steady-state conditional
FCS for the SQD-QPC system  \cite{Sukhorukov07,Jordan04}.

One of the purposes of this paper is to provide a connection with, and a
unified picture of, the quantum trajectory, the
(partially reduced) 
number-resolved master-equation and the unconditional (reduced)
master-equation approaches. We show that the master equations for the reduced or
partially  reduced density matrix can be simply obtained when an
average or partial average is taken on the conditional, stochastic
density matrix (or quantum trajectories) over the possible outcomes of
the measurements \cite{Goan01a,Goan01b,Goan03}.

Another purpose of this paper is to investigate the effect of QPC shot
noise on the conditional FCS of the SQD-QPC system as well as to 
develop an efficient and
systematic way to calculate the conditional FCS for more complex
systems of
interacting nanoscale conductors. 
Our investigation goes beyond the analysis presented in
Ref.~\onlinecite{Sukhorukov07}.
In Ref.~\onlinecite{Sukhorukov07}, 
the number-resolved population
master (rate) equation for bistable system was first transformed 
into the counting field (inverse Fourier transform) space  
and then the eigenvalue with the smallest
absolute real part (or maximum eigenvalue) in the matrix of the transformed
master equation was found analytically.
To evaluate the integral in partial or mixed
Fourier transform space  
analytically with the analytic form of the 
eigenvalue 
to obtain the conditional steady-state current moment (cumulant) generating
function, 
a further approximation to neglect the QPC shot noise was made\cite{Sukhorukov07}. 
For the experimental parameters used in Ref.~\onlinecite{Sukhorukov07},
the QPC shot noise 
as compared to the noise contribution of the random telegraph signal
in the QPC current trace 
induced by random electrons tunneling on and off the QD 
is indeed small and can be neglected.
On the other hand, for the 
parameter regimes where the QPC shot noise cannot be ignored,
obtaining analytical expressions for the conditional steady-state current
moments or cumulants is very difficult. 
Furthermore, for more complicated interacting nanoscale conductors
with the dimension of the matrix equation of the master equation
growing up, to find analytical solution of the maximum eigenvalue becomes 
very hard, not to mention to obtain the analytical forms of the conditional
steady-state current moment or current cumulant generating function.
Besides,
direct numerical evaluation of the conditional cumulant
generation function 
in the same way as in Ref.~\onlinecite{Sukhorukov07}
and then taking partial derivatives to obtain conditional cumulants
are quite numerically unstable. 
In these cases, 
the quantum trajectory approach may  give the conditional states
or conditional current cumulants by simultaneously simulating  
an ensemble of current outcomes and corresponding quantum
trajectories, and then categorizing and averaging the current outcomes
of one system (e.g., the QD system) 
for each of the observed average current value in the other system
(e.g., the QPC).
However, in some parameter regimes where 
the probabilities to observe the average QPC current in
certain values are very small,  
it is then computationally expensive to simulate and map
out the conditional current cumulant in the whole parameter space of
the average QPC current
by the quantum trajectory method as an extremely large number of 
trajectories are required to have enough statistical samples in those very low
probability domains. Thus developing a method to evaluate the
conditional counting statistics directly and 
effectively for more complex systems and for 
a wide range of parameter space is desirable. 
It is one of the 
aims of the paper to develop such an efficient method.

The paper is organized as follows. 
In Sec.~\ref{sec:Model}, we introduced the model and
Hamiltonian of the QD-QPC system that will be considered. In
Sec.~\ref{sec:UME}, we present the 
unconditional master equation for the reduced density matrix of the
QD system. We then derive the conditional, stochastic master equation (or quantum
trajectory equation) that mimics the
dynamics of the QD system conditioned on the observed random outcomes
in Sec.~\ref{sec:Qtrajectory}.
Then the number-resolved master equation and its inverse Fourier transform in
the counting field space are discussed in Sec.~\ref{sec:NRME}. The
procedure to calculate the 
unconditional and conditional FCS are described in terms of generating
functional approach in Sec.~\ref{sec:CS}. Here we also introduce our
efficient method to calculate the moments and cumulants 
 of the conditional FCS.
Section \ref{sec:Results} is devoted to the presentation and
discussion of the results we obtain. 
Specifically, 
we provide a thorough analysis using the method of
Ref.~\onlinecite{Sukhorukov07}, the 
quantum trajectory theory
and the efficient method we develop to simulate and calculate 
the conditional current and noise of the SQD-QPC and
DQD-QPC systems. We also investigate how the QPC shot noise affects the
conditional QD current cumulants.  
Finally, a short conclusion is given in Sec.~\ref{sec:Conclusion}. 
The detailed procedure of the semiempirical method
used to count the number of tunneling electrons through the QD system 
in each random current trace of 
quantum trajectories 
is described in Appendix \ref{App:counting_method}.


\section{Quantum-dot system measured by a QPC}
\label{sec:Model}

\begin{figure}
\includegraphics[width=8cm]{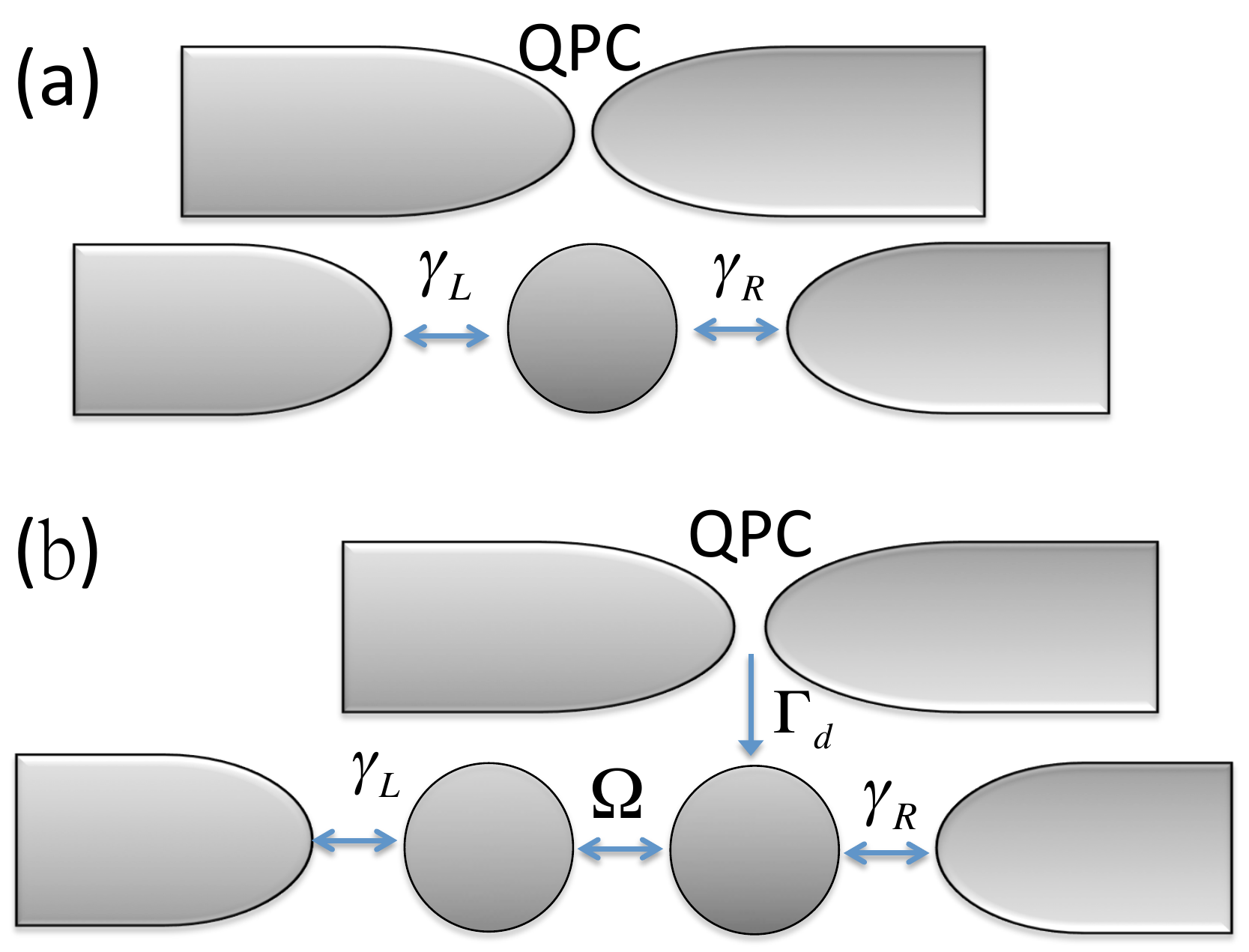}

\caption{Schematic illustration of 
(a) SQD and (b) coherently coupled DQD's connected
to two Fermi reservoirs (left and right leads) by tunnel junctions,
measured by a charge-sensitive QPC detector.
Electrons tunneling through the QD modulate the tunneling current
through the QPC. }
\label{fig:QD_setup}
\end{figure}

We consider a system consisting of either a SQD
[see Fig.~\ref{fig:QD_setup}(a)]
or coherently coupled
DQD's [Fig.~\ref{fig:QD_setup}(b)] measured by a QPC
\cite{Gurvitz97,Schoell06,Gustavsson06,Sukhorukov07,Gustavsson09}.
The QD system is connected to two leads (reservoirs)
 biased so that electrons can tunnel onto the SQD (onto the left dot of the
 coherently coupled DQD's) from the
left lead and off the SQD (off the right dot of the DQD's) onto the right lead.
The QPC serves as a sensitive electrometer 
since its tunneling barrier
can be modulated by the charge on a nearby QD.
In our setup, as the electron
moves into the SQD (the right dot of the DQD's), it changes the tunnel
barrier of the nearby QPC. In this way the modulated current through the QPC
can be used to continuously monitor the occupation of the QD. We
will follow the treatment given in Ref.~\onlinecite{Goan01a,Goan01b,Goan03} to
describe the dynamics of the system.

The Hamiltonian for the QD system coupling to the QPC  can be
written as 
\begin{equation}
{\cal H}= {\cal H}_{QD}+{\cal H}_{QPC}+{\cal H}_{coup}
\end{equation}
where
\begin{eqnarray}
{\cal H}_{QD}&=&H_S + \sum_k \hbar \omega_{Lk}
a_{Lk}^\dagger a_{Lk} + \hbar \omega_{Rk} a_{Rk}^\dagger a_{Rk}
\nonumber \\
&&+ \sum_k t_{Lk} a_{Lk} c_i^\dagger+ t_{Rk} a_{Rk} c_j^\dagger+ h.c
\label{HCQD} \\
{\cal H}_{QPC}&=&\hbar \sum_k \left(\omega_{sk} a_{sk}^\dagger a_{sk}
+\omega_k^R a_{dk}^\dagger a_{dk}\right) \nonumber\\
&&+ \sum_{k,q} \left(T_{kq}a_{sk}^\dagger a_{dq}+
T^*_{qk}a_{dq}^\dagger a_{sk} \right),
\label{HPC} \\
{\cal H}_{coup}&=&\sum_{k,q} c_j^\dagger c_j
\left(\chi_{kq}a_{sk}^\dagger a_{dq}
+ \chi^*_{qk}a_{dq}^\dagger a_{sk} \right).
\label{Hcoup}
\end{eqnarray}
where ${\cal H}_{QD}$ here is the Hamiltonian for the QD
system consisting of the left lead, right lead and the central part system
and the tunneling between them.  The symbols $a_{Lk}$,
$a_{Rk}$ and $\hbar\omega_{Lk}$, $\hbar\omega_{Rk}$ are respectively
the electron annihilation operators and energies for the left and
right reservoir states for the QD system at wave number $k$. 
For a SQD system, we have 
the indices $i=j=2$ in  ${\cal H}_{QD}$ and the Hamiltonian of the central part
system is just 
\begin{equation}
  \label{eq:SQDH}
 H_S = \hbar \omega_2 c_2^\dagger c_2,
\end{equation}
and for a DQD system, we have $i=1$, $j=2$ in  ${\cal H}_{QD}$ and 
\begin{equation}
  \label{eq:DQDH}
H_S= \hbar \omega_1 c_1^{\dagger}c_1+\hbar \omega_2 c_2^{\dagger}c_2+\hbar \Omega(c_1^{\dagger}c_2+c_2^{\dagger}c_1).
\end{equation}
Here $c_j$ ($c_j^\dagger$) and $\hbar\omega_j$ represent
the electron annihilation (creation) operator and energy for a
single electron state in dot $j$, respectively.
In other words, dot 2 denotes the central QD in the SQD
system, and dot 1 and dot 2 stand for the left dot and right dot,
respectively, in the DQD system. 
The tunneling Hamiltonian for the QPC detector is represented by
${\cal H}_{QPC}$.  Similarly, $a_{sk}$,
$a_{dk}$ and $\hbar\omega_{sk}$, $\hbar\omega_{dk}$ are respectively
the electron annihilation operators and energies for the source and
drain reservoir states for the QPC at wave number $k$. ${\cal
H}_{coup}$ [Eq. (\ref{Hcoup})] describes the interaction between the
QPC detector and dot $j=2$. When the electron is located in dot $j=2$, the
effective tunneling amplitude of the QPC detector changes from
$T_{kq}\rightarrow T_{kq}+\chi_{kq}$.
We investigate here a simpler case of 
electrons transport through the DQD-QPC system 
in which the
QPC couples only to the right dot (dot 2) of the DQD system
\cite{Gurvitz97,Schoell06,Gustavsson09,Marcus04}
to illustrate the usage of our method and discuss
the effects of QPC shot noise and interdot coupling on the conditional
current cumulants.
Our approach can be straightforwardly generalized to the case where  
the QPC couples to both dots with different coupling strengths
\cite{Fujisawa06,Fricke07,Gustavsson09,key-36,key-38,key-39}.

\section{Unconditional master equation}
\label{sec:UME}
By following the treatment in Refs.~\onlinecite{Gurvitz97,Goan01a,Goan01b},
the (unconditional) zero-temperature, Markovian master equation of
the reduced density matrix
for the
quantum dot (QD) system can be obtained as:
\begin{eqnarray}
\dot{\rho}(t)&=&-\frac{i}{\hbar}[H_S,\rho(t)]
+\gamma_L{\cal D}[c_i^\dagger]\rho(t)+\gamma_R{\cal D}[c_j]\rho(t)\nonumber\\
&&+{\cal D}[{\cal T}+{\cal X} n_j]\rho(t)
\label{masterEq}
\\
&\equiv&{\cal L} \rho(t),
\label{Liouvillian}
\end{eqnarray}
where $n_j=c_j^\dagger c_j$ is the occupation number operator of 
dot $j$ measured by the QPC. 
In Eq.~(\ref{masterEq}) and the rest of the paper, the Hamiltonian of
the QD system $H_s$ takes the form of Eq.~(\ref{eq:SQDH}) for the SQD
system and  Eq.~(\ref{eq:DQDH}) for the DQD system, and the subscripts
$i=j=2$ for the SQD
system and $i=1$ and $j=2$ for the DQD system. 
The parameters ${\cal T}$ and ${\cal X}$ are given by $D=|{\cal
T}|^2= 2\pi  |T_{00}|^2 g_s g_d V/\hbar$, and $D'=|{\cal T}+{\cal
X}|^2=2\pi  |T_{00}+\chi_{00}|^2 g_s g_d V/\hbar$. Here $D$ and
$D'$ are the average electron tunneling rates through the QPC
barrier without and with the presence of the electron in dot $j=2$
respectively, $eV=\mu_s-\mu_d$ is the external bias applied across
the QPC ($\mu_s$ and $\mu_d$ stand for the chemical potentials in
the source and drain reservoirs, respectively), $T_{00}$ and $\chi_{00}$ are
energy-independent tunneling amplitudes near the average chemical
potential, and $g_s$ and $g_d$ are the energy-independent density of
states for the source and drain reservoirs.   $\gamma_L$ and
$\gamma_R$ are the tunneling rates from the left lead 
 to the QD system  and from the 
QD system to the right lead, respectively.
In Eq.~(\ref{masterEq}), the superoperator ${\cal D}$ is defined as:
\begin{equation}
{\cal D}[B]\rho={\cal J}[B]\rho - {\cal A}[B]\rho,
\label{defcalD}
\end{equation}
where ${\cal J}[B]\rho = B \rho B^\dagger, {\cal A}[B]\rho =
(B^\dagger B \rho +\rho B^\dagger B)/2$. Finally, Eq.\
(\ref{Liouvillian}) defines the Liouvillian operator ${\cal L}$.

The conditional dynamics is quite different from its unconditional
counterpart. For example, the unconditional dynamics
of the number of electrons on the SQD system
follows immediately from Eqs.~(\ref{masterEq}) and (\ref{eq:SQDH})
as 
\begin{equation}
\frac{d\langle n_2\rangle (t)}{dt} = \gamma_L[1-\langle n_2\rangle(t)]-\gamma_R\langle n_2\rangle(t),
\end{equation}
where $\left\langle n_2\right\rangle (t)=Tr[c_2^{\dagger}c_2\rho(t)]$.
Clearly the average current through the SQD does not depend at all on
the current through the QPC in this model. This is because the
Hamiltonian describing the interaction between the SQD and the QPC
commutes with the number operator $n_2$. However if we ask for the
conditional dynamics of the SQD {\em given} an observed averaged
current in time $t$ or {\em given} an instantaneous current in time $dt$ 
through the QPC, we need a different equation or even a stochastic
equation for $\langle n_2\rangle_c$.

\section{Conditional master equation and quantum trajectories}
\label{sec:Qtrajectory}

There are two classical stochastic currents through this system: the
current, $I(t)$, through the QPC and the current, $J(t)$, through
the QD. Equation (\ref{masterEq}) describes the time
evolution of the reduced 
density matrix when these classical stochastic processes are averaged
over. To make contact with a single realization of the random outcomes
of the measurement
records and study the stochastic evolution of the QD state,
conditioned on a particular measurement realization, we need the conditional
master equation. We first define the relevant point processes that
are the source of the classically observed stochastic currents. 

We specify the quantum jump conditional dynamics through the QPC by
defining the point processes \cite{Goan01a,Goan01b,Goan03,Goan04}: 
\begin{eqnarray}
[dN_{c}(t)]^{2} & =&dN_{c}(t),\\
E[dN_{c}(t)] & =&\zeta Tr[\widetilde{\rho}_{1c}(t+dt)] \nonumber\\
 & =&\zeta[D+(D'-D)\left\langle n_2\right\rangle _{c}(t)]dt \nonumber\\
&=&\zeta\mathcal{P}_{1c}(t)dt,
\label{ineffdNav}
\end{eqnarray}
where $dN_{c}(t)$ is a stochastic point process which represents
the number (either zero or one) of tunneling events in the QPC
seen in an infinitesimal
time $dt$,
\begin{equation}
\widetilde{\rho}_{1c}(t+dt)=\mathcal{J}[T+\chi n_2]\rho_{c}(t)dt
\end{equation}
is the unnormalized density matrix\cite{Goan01a,Goan01b} given the result
of an electron tunneling through the QPC barrier at the end of the
time interval $[t,t+dt)$,
$\langle
n_2\rangle_c(t)={\rm Tr}[n_2\rho_c(t)]$, 
\begin{equation}
{\cal P}_{1c}(t)= D+(D'-D)\langle n_2\rangle_c(t),
\label{P1c}
\end{equation}
and $E[Y]$ denotes an ensemble average of a classical stochastic
process $Y$. The subscript $c$ indicates that the quantity to which
it is attached is conditioned on previous observations of the the
occurrences (detection records) of the electrons tunneling through
the QPC barrier in the infinitesimal time $dt$ in the past.
The factor $\zeta\leq1$ represents the
fraction of tunneling events which are actually registered by the
circuit containing the QPC detector. The value $\zeta=1$ then corresponds
to a perfect detector or efficient measurement. By using the fact that
current through the QPC is $I(t) 
= e\, {dN(t)}/{dt}$, Eq.~(\ref{ineffdNav}) with $\zeta=1$
states that the average current is $eD$ when the dot is empty, and
is $eD'$ when the dot is occupied.

Similarly, we can specify the quantum jump conditional dynamics through the
QD system by defining two stochastic point processes $dM_{Lc}(t)$ and
$dM_{Rc}(t)$ which represent, respectively,
the numbers (either zeros or ones) of tunneling events from the left
lead to dot $i$ and from dot $j$ to the right lead
seen in an infinitesimal
time $dt$:
\cite{Goan01a,Goan01b,Goan03,Goan04}: 
\begin{eqnarray}
[dM_{Lc}(t)]^{2} & = & dM_{Lc}(t), 
\quad [dM_{Rc}(t)]^2  =  dM_{Rc}(t),\\
E[dM_{Lc}(t)] & = & \gamma_{L}\langle c_ic_i^{\dagger}\rangle_c
(t)dt=\gamma_{L}[1-\left\langle n_i\right\rangle_{c}(t)]dt,\\
E[dM_{Rc}(t)] & = & \gamma_{R}\langle c_j^{\dagger}c_j\rangle_c
(t)dt=\gamma_{R}\langle n_j\rangle_{c}(t)dt,
\end{eqnarray}
where $\left\langle n_j\right\rangle _{c}(t)=Tr[c_j^{\dagger}c_j\rho_c(t)]$.

Unraveling both the QPC and the QD equations, we write the
conditional master equation at zero temperature as:
\begin{eqnarray}
d \rho_c (t)&=&
 dM_{Lc} \bigg[ \frac{\mathcal{J}[c^\dagger]}{1-\langle n_i\rangle_c(t)}
- 1 \bigg] \rho_c (t) \nonumber\\
&&
+ dM_{Rc} \bigg[ \frac{\mathcal{J}[c]}{\langle n_j\rangle_c(t)} -
1 \bigg] \rho_c(t) \nonumber \\
&&- dt\{\gamma_L \mathcal{A}[c_i^\dagger] \rho_c(t)+ \gamma_R
\mathcal{A}[c_j] \rho_c(t)\nonumber\\
&&
\quad
- \gamma_L[1-\langle n_i\rangle_c(t)]\rho_c(t) - \gamma_R \langle n_j\rangle_c(t)\rho_c(t)\} \nonumber\\
&&+ dN_c \bigg[ \frac{\mathcal{J}[\mathcal{T}+\chi
n_2]}{\mathcal{P}_{c}(t)} - 1 \bigg] \rho_c (t)\nonumber \\
&& +\, dt  \{ -({i}/{\hbar})[H_S,\rho_c(t)]-{\cal A}[{\cal T}+{\cal X} n]\rho_c(t)\nonumber \\
&&
+(1-\zeta){\cal J}[{\cal T}+{\cal X} n] \rho_c(t)
 +\zeta \,  {\cal P}_{1c}(t) \rho_c(t) \} 
\label{eq:Cond_Unravel}
\end{eqnarray}

We now focus on the conditional dynamics of the QD as the
QPC current, $I(t)$, is continuously monitored. In the experiment the observed
values of the random telegraph process are not fixed at the average
values, $D$,$D'$, but are themselves stochastic processes as electrons
tunnel through the QPC. We average over the jump process onto
and off the QD.
The stochastic quantum-jump master equation of the density matrix
operator, conditioned on the observed event in QPC current in the case
of inefficient 
measurement in time $dt$ can be obtained as, 
\begin{eqnarray}
d\rho_c(t)
&=&dN_c(t)\left [\frac{{\cal J}[{\cal T}+{\cal X} n]}
{{\cal P}_{1c}(t)}
-1\right ]\rho_c(t)
\nonumber \\
&& +\, dt  \{ -({i}/{\hbar})[H_S,\rho_c(t)]-{\cal A}[{\cal T}+{\cal X} n]\rho_c(t)\nonumber \\
&&\quad\quad+(1-\zeta){\cal J}[{\cal T}+{\cal X} n] \rho_c(t)
 +\zeta \,  {\cal P}_{1c}(t) \rho_c(t) \nonumber \\
&&\quad\quad+\gamma_L{\cal D}[c_i^\dagger]\rho_c(t) +\gamma_R{\cal D}[c_j]\rho_c(t)\}.
\label{condmasterEq}
\end{eqnarray}
In the quantum-jump case,
in which individual electron QPC tunneling current events can be
distinguished, the QD system state [see Eq.\ (\ref{condmasterEq})]
undergoes a finite evolution (a
{\em quantum jump}) when there is
a detection result [$dN_c(t)=1$]
at randomly determined times (conditionally Poisson
distributed).

As Fig.~1(c) of Ref.~\onlinecite{Sukhorukov07} suggests, the
current through the QPC could be quite 
large and while we may be able to resolve the random telegraph signal 
jumps between the
two average values, $D$ and $D'$, we may not have sufficient bandwidth
in the circuit to resolve the jump events $dN(t)$ through the QPC.
The individual tunnel events through the QPC are too rapid to be resolved
in the external circuit, resulting in a process more like a white
noise stochastic process. This leads us to make the diffusive approximation
to the quantum-jump stochastic master equation for describing the
conditional QPC current dynamics. We now replace the quantum-jump
master equation for the QPC with the quantum diffusion stochastic
master equation. In this case, the total number of electrons that
tunnel through the QPC in a time $\delta t$, large compared to the
inverse of the
jump rate, but small compared to the typical circuit response time,
is considered as a continuous
diffusive variable 
satisfying a Gaussian white noise distribution \cite{Goan01a,Goan01b}: 
\begin{equation}
\delta N(t)=\{\zeta \,  |{\cal T}|^2 [1+2\, \epsilon\,
\cos\theta\,\langle n_2\rangle_c(t)] +\sqrt{\zeta\, } \, |{\cal T}|
\xi(t)\} \delta t, \label{deltaN}
\end{equation}
where $\epsilon=(|{\cal X}|/|{\cal T}|)\ll 1$, $\theta$ is the
relative phase between ${\cal X}$ and ${\cal T}$, and $\xi(t)$ is a
Gaussian white noise characterized by
\begin{equation}
E[\xi(t)]=0, \quad E[\xi(t)\xi(t')]=\delta(t-t'). \label{xi}
\end{equation}
Here $E$ denotes an ensemble average. 
In stochastic calculus, $\xi(t)dt=dW(t)$ is known as the infinitesimal
Wiener increment.
In obtaining Eq.\
(\ref{deltaN}), we have assumed that $2|{\cal T}||{\cal X}|\,
\cos\theta \gg |{\cal X}|^2$. Hence, for the quantum-diffusive
equations obtained later, we should regard, to the order of
magnitude, that $|\cos\theta|\sim O(1)\gg \epsilon=(|{\cal
X}|/|{\cal T}|)$ and $|\sin\theta|\sim O(\epsilon)\ll 1$.

By taking the diffusive limit on the QPC, 
the quantum-diffusive conditional master equation for the case of
inefficient measurements can be found as:
\begin{eqnarray}
\dot{\rho}_c(t) &&=-\frac{i}{\hbar}[H_S,\rho_c(t)]\nonumber \\
&&+\gamma_L{\cal D}[c_i^\dagger]\rho_c(t)+\gamma_R{\cal D}[c_j]\rho_c (t)+{\cal
D}[{\cal T}+{\cal X}n_2]\rho_c(t)
\nonumber \\
&&+\xi(t)\frac{\sqrt{\zeta\, } \, }{|{\cal T}|} [{\cal T}^* {\cal
X}\,n_2\rho_c(t)+ {\cal X}^* {\cal T}\rho_c(t)n_2 \nonumber
\\
&&-2\,{\rm Re}({\cal T}^* {\cal X})\langle n_2\rangle_c(t)\rho_c(t)].
\label{diffusivemasterEq}
\end{eqnarray}

We will now make the simplifying assumption that $\theta=0$. In that
case ${\cal T}$ and ${\cal \chi}$ are real and $D=|{\cal T}|^2,\
D'=|{\cal T}+{\cal \chi}|^2$. This corresponds to $D'>D$ as in the
experiment of Ref.~\onlinecite{Sukhorukov07}.
The conditional current through the QPC, $I_c(t)=e\delta N(t)/\delta
t$, conditioned on the dot occupation,  satisfies the stochastic
differential equation 
\begin{equation}
I_c(t)=e\zeta D[1-2\epsilon\langle n_2\rangle_c(t)]+e\sqrt{\zeta D}\xi(t)
\label{I_c}
\end{equation}
with
\begin{equation} 
\epsilon=1-\sqrt{\frac{D'}{D}}.      
\end{equation}

We can now find from Eq.~(\ref{diffusivemasterEq}) the conditional
dynamics of the dot occupation 
conditioned on the observed instantaneous QPC current in time $dt$. 
For the SQD-QPC system, we have
\begin{eqnarray}
\frac{d\langle n_2\rangle_c(t)}{dt}&=& \gamma_L[1-\langle n_2\rangle_c(t)
]-\gamma_R\langle n_2\rangle_c(t)\nonumber \\
&&-2\chi\sqrt{\zeta}[1-\langle
n_2\rangle_c(t)]\langle n_2\rangle_c(t)\xi(t)
\label{dn_c}
\end{eqnarray}
Note that the noise ``turns off'' when the dot (dot 2) is either occupied or
empty. This can be understood if we regard the QPC current as a
measurement of the dot occupation. Suppose that  $\gamma_L\neq 0$, and 
$\gamma_R=0$, in which case an electron will eventually tunnel onto the
dot. The QPC current must eventually revel this fact, as the current
through the QPC will increase. After a small interval of time we will
be confident that this is a real effect and not a random fluctuation
and the conditional mean $\langle n_2\rangle_c$ becomes locked on unity
with no further fluctuation. A parallel argument can be made in the
case that $\gamma_L= 0,\ \gamma_R\neq 0$. We thus see that this
feature of the noise is a reflection of the fact that monitoring the
QPC current gives us information on the state of the QD.

Similarly, we obtain
from Eq.~(\ref{diffusivemasterEq}) the equations of motion to
determine the DQD coherence and occupations
conditioned on the observed instantaneous QPC current in time $dt$ as
\begin{eqnarray}
\frac{d\langle n_{1}\rangle _{c}}{dt} & =&\gamma_{L}(1-\langle
n_{1}\rangle _{c})-i\Omega(\langle c_{2}^{\dagger}c_{1}\rangle
_{c}-\langle c_{1}^{\dagger}c_{2}\rangle _{c}) \nonumber\\
 && \;+2\chi\sqrt{\zeta}\xi(t)(\langle n_{1}n_{2}\rangle _{c}-\langle
 n_{1}\rangle _{c}\langle n_{2}\rangle _{c})
\label{eq:n1c}
\\
\frac{d\langle n_{2}\rangle _{c}}{dt} & =&-\gamma_{R}\langle n_{2}\rangle _{c}-i\Omega(\langle c_{1}^{\dagger}c_{2}\rangle _{c}-\langle c_{2}^{\dagger}c_{1}\rangle _{c})\nonumber\\
 && \;+2\chi\sqrt{\zeta}\xi(t)(\langle n_{2}\rangle _{c}-\langle n_{2}\rangle _{c}^{2})\\
\frac{d\langle c_{1}^{\dagger}c_{2}\rangle _{c}}{dt} &
=&-\frac{\gamma_{L}+\gamma_{R}+\chi^{2}}{2}\langle
c_{1}^{\dagger}c_{2}\rangle _{c}+i\Omega(\langle n_{2}\rangle
_{c}-\langle n_{1}\rangle _{c})\nonumber\\
 && \;
+2\chi\sqrt{\zeta}\xi(t)(\frac{1}{2}-\langle n_{2}\rangle _{c})\langle c_{1}^{\dagger}c_{2}\rangle _{c}\\
\frac{d\langle c_{2}^{\dagger}c_{1}\rangle _{c}}{dt} & =&-\frac{\gamma_{L}+\gamma_{R}+\chi^{2}}{2}\langle c_{2}^{\dagger}c_{1}\rangle _{c}+i\Omega(\langle n_{1}\rangle _{c}-\langle n_{2}\rangle _{c})\nonumber\\
 && \;
+2\chi\sqrt{\zeta}\xi(t)(\frac{1}{2}-\langle n_{2}\rangle _{c})\langle c_{2}^{\dagger}c_{1}\rangle _{c}\\
\frac{d\langle n_{1}n_{2}\rangle _{c}}{dt} &
=&-(\gamma_{L}+\gamma_{R})\langle n_{1}n_{2}\rangle
_{c}+\gamma_{L}\langle n_{2}\rangle _{c} \nonumber\\
 && \:+2\chi\sqrt{\zeta}\xi(t)(1-\langle n_{2}\rangle _{c})\langle
 n_{1}n_{2}\rangle _{c}.
\label{eq:n1n2c}
\end{eqnarray}
It is understood that all the conditional quantum average quantities
in Eqs.~(\ref{eq:n1c})--(\ref{eq:n1n2c}) carry time
dependence, i.e.,  
$\langle\cdots\rangle_c\equiv\langle\cdots\rangle_c(t)$. 

\begin{figure}
\includegraphics[width=8cm]{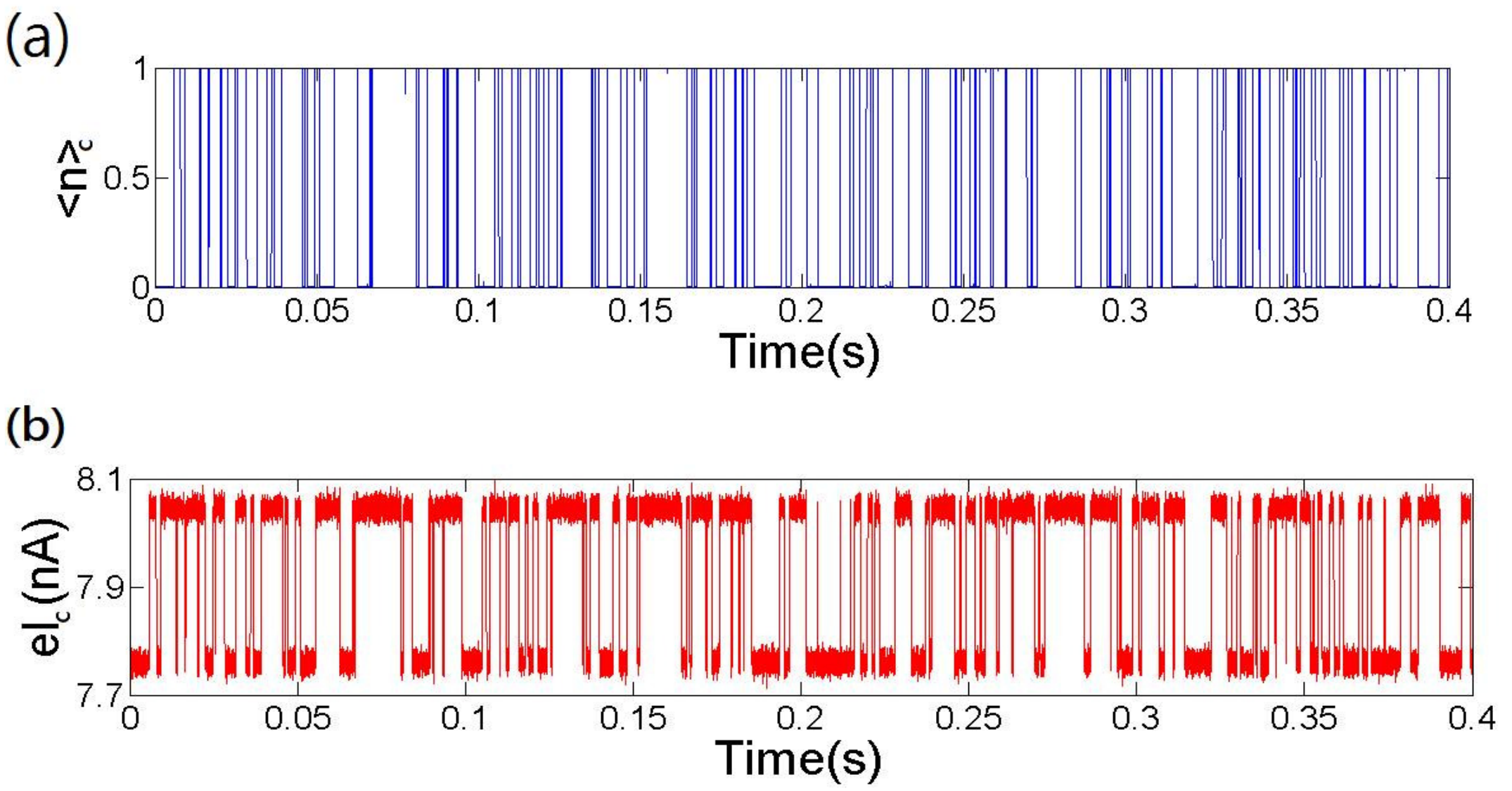}

\caption{Simulation of (a) the 
conditional expectation value of the electron occupation number $\left\langle
  n_2\right\rangle_c $ and (b) the QPC current. 
The QPC current is taken with bandwidth
of 100 kHz and through a Butterworth filter with eight order and cutoff
frequency 4 kHz as in the experiment of Ref.~\onlinecite{Sukhorukov07}
with parameters
$(D,\,
D',\,\gamma_{L},\gamma_{R})=(4.85\times10^{10},\,5.03\times10^{10},160,586)$ Hz.
\label{fig:Fig2}
}
\end{figure}

The quantum trajectory theory provides us with full information of the
statistical 
properties about the measured system as that of
an experimentalist who actually performs 
a time-resolved transport experiment. 
One can use Eq.~(\ref{dn_c}) for the SQD system and 
Eqs.~(\ref{eq:n1c})--(\ref{eq:n1n2c}) for the DQD system 
to calculate the conditional charge occupation 
number $\langle n_2\rangle _{c}(t)$ and then use Eq.~(\ref{I_c}) to mimic 
the measured QPC current record $I_c(t)$ continuously in time  
 in a single run of a realistic experiment.
We show in Fig.~\ref{fig:Fig2}(a) a typical realization of the
trajectories of $\langle n_2\rangle _{c}(t)$ 
 for the SQD-QPC system 
obtained by the quantum trajectory theory 
and  in Fig.~\ref{fig:Fig2}(b) its corresponding conditional QPC
current $I_c(t)$ 
taking into account the detection bandwidth in experiment \cite{Sukhorukov07}. 
The simulated QPC current shows random switchings
between two average currents, which  
correspond to the single-electron tunneling onto and off the QD.
It indeed resembles the typical measured QPC current shown in Fig.~1(c) of
 Ref.~\onlinecite{Sukhorukov07}. 
Simulating a great amount of trajectories by many different random
realizations of $\xi(t)$, one is able to calculate all the statistical
transport quantities of the QD systems, such as the conditional counting statistics.   
One can obtain the time average QPC
current $I$ in time $t$ by integrating the instantaneous QPC current $I_c(t)$
and acquire the average QD current $J$ conditional on QPC current $I$ in
time $t$ in its corresponding $\langle n_2 \rangle_c(t)$ trajectory (see
Sec.~\ref{sec:CCSbyQT} for details).  
We simulate a great amount of trajectories
and sort $J$ by different $I$, and use the formulas of the
conditional  moments and 
cumulants to obtain the conditional counting statistics from the data collected from these trajectories.

It is easy to see that the ensemble
average evolution of the conditional master equation, 
Eq.\ (\ref{diffusivemasterEq}), 
reproduces the unconditional master equation
(\ref{masterEq}) by simply eliminating the white noise term using
Eq.\ (\ref{xi}).
Similarly, averaging Eq.\ (\ref{condmasterEq}) over
the observed stochastic process, by
setting $E[dN_c(t)]$ equal to its expected value Eq.\ (\ref{ineffdNav}),
gives the unconditional, deterministic
master equation (\ref{masterEq}).
It is also easy to verify that for zero efficiency $\zeta=0$
[i.e., also $dN_c(t)=0$],
the conditional equations (\ref{condmasterEq}) and (\ref{diffusivemasterEq})
reduce to the unconditional one, (\ref{masterEq}).
That is, the effect of averaging over all possible
measurement records is equivalent to the effect of completely ignoring
the detection records or the effect of no detection results being available.

\section{Number-resolved master equation}
\label{sec:NRME}
To study the current cumulants of one conductor (e.g., the QD system)
conditioned on the average current of the other conductor (e.g. the QPC),
we turn to the 
number-resolved master equation \cite{Gurvitz97, Korotkov01,
Shnirman98,Makhlin00} or the master equation for the partially
reduced density matrix \cite{Goan03,Goan04} of the joint QD and QPC
system. If $N$ electrons have tunneled 
through the QPC and $M$ electrons have tunneled through the right
junction of the QD at time $t+dt$, then the accumulated
number of electrons in the drain of the QPC at the earlier time $t$,
due to the contribution of the {\em jump} term of the QPC, should be
$(N-1)$ for $M$ electron in the drain of the QD, and it
should be $(M-1)$ in the drain (right lead) of the QD system due to the
contribution of the {\em jump} term of the QD for $N$
electron in the drain of the QPC \cite{Goan03,Goan04}. Hence,
after writing out the number dependence $N$, $(N-1)$, $M$, or $M-1$
explicitly for the density matrix in Eq.\ (\ref{masterEq}), we
obtain the master equation for the ``partially'' reduced density
matrix as:
\begin{eqnarray}
\dot{\rho}(N,M,t)&=&
-({i}/{\hbar})[H_S,\rho_c(N,M,t)]\nonumber \\
&&+\zeta {\cal J}[{\cal T}+{\cal X} n]\rho(N-1,M,t)\nonumber \\
&&+(1-\zeta){\cal J}[{\cal T}+{\cal X} n] \rho(N,M,t)\nonumber \\
&&-{\cal A}[{\cal T}+{\cal X} n]\rho(N,M,t) \nonumber \\
&&+\gamma_L{\cal D}[c_i^\dagger]\rho(N,M,t)
-\gamma_R{\cal A}[c_j]\rho(N,M,t)\nonumber \\
&&+ \gamma_R{\cal J}[c_j]\rho(N,M-1,t) 
\label{masterEqN}
\end{eqnarray}
If the sum over all possible
values of $N$ and $M$ is taken on the ``partially'' reduced density
matrix [i.e., $\rho(t)=\sum_{N,M} \rho(N,M,t)$], Eq.\
(\ref{masterEqN}) then reduces to Eq.\ (\ref{masterEq}). 

For
simplicity, in the following we set the QPC detection efficiency $\zeta=1$ 
corresponding to perfect detections or efficient measurements. 
We deal with the case of the SQD-QPC system first.  After
evaluating Eq.\ (\ref{masterEqN}) in the occupation number basis
$|0\rangle$ and $|1\rangle$ of the SQD, we obtain the rate
equations as:
\begin{eqnarray}
\dot\rho_{00}(N,M,t) &=& |{\cal T}|^2\rho_{00}(N-1,M,t)-|{\cal
T}|^2\rho_{00}(N,M,t)
\nonumber \\
&&\hspace*{-0.7cm}-\gamma_L\rho_{00}(N,M,t)+\gamma_R\rho_{11}(N,M-1,t)
\;,
\label{rateEq0}\\
\dot\rho_{11}(N,M,t) &=& |{\cal T}+{\cal
X}|^2\rho_{11}(N-1,M,t)\nonumber \\
&&-|{\cal T} +{\cal X}|^2\rho_{11}(N,M,t)
\nonumber \\
&&+\gamma_L\rho_{00}(N,M,t)-\gamma_R\rho_{11}(N,M,t)
\;,
\label{rateEq1}
\end{eqnarray}
where $\rho_{aa}=\langle a|\rho |a\rangle$ with $a=0,1$ referring to the
QD occupation number states.

To obtain the solution of $\rho_{aa}(N,M,t)$ with $a=0,1$ in the
number-resolved or the ``partially'' reduced density matrix
approach, we can first apply a two-dimensional Fourier transform
(to the counting field space)
\cite{Shnirman98,Makhlin00,Goan03,Goan04}
\begin{equation}
\rho_{aa}(k,q,t)=\sum_{N,M} e^{ikN+iqM}\rho_{aa}(N,M,t) \label{rhokq}
\end{equation}
to Eqs.\ (\ref{rateEq0}) and (\ref{rateEq1}) since these equations are
translationally invariant in $N$ and $M$ space. So after the Fourier
transform, we obtain from Eqs.\ (\ref{rateEq0}) and (\ref{rateEq1})
that
\begin{equation}
  \label{eq:LME}
\frac{d \rho(k,q,t)}{dt}={\cal L}(k,q)\rho(k,q,t),
\end{equation}
where 
\begin{eqnarray}
{\rho(k,q,t)}&=&\left(\begin{array}{c}\rho_{00}(k,q,t)\\
\rho_{11}(k,q,t)\end{array}\right),\\
{\cal L}(k,q)&=& 
\left(\begin{array}{cc}
D(e^{ik}-1)-\gamma_L & \gamma_R e^{iq}\\
\gamma_L & D'(e^{ik}-1)-\gamma_R
\end{array}\right) \;.
\label{Lkq}
\end{eqnarray}
We note here again that we have set ${\cal T}$ and ${\cal \chi}$ to be
real and 
their relative phase angle $\theta=0$ so that $D=|{\cal T}|^2,\
D'=|{\cal T}+{\cal \chi}|^2$.

Similarly for the case of coherently coupled  DQD's measured by a 
QPC [see Fig.~\ref{fig:QD_setup}(b)],  
the number-resolved master equation in the Fourier space (counting
field space)
can also be written in the form of Eq.~(\ref{eq:LME}) with 
$\mathcal{L}(k,q)$ given in matrix form as 
\begin{equation}
\begin{array}{l}
\mathcal{L}(k,q)=\\
\left(\begin{array}{cccccc}
f(k)-\gamma_{L} & 0 & \gamma_{R}e^{iq} & 0 & 0 & 0\\
\gamma_{L} & f(k) & 0 & \gamma_{R}e^{iq} & 0 & 2\Omega\\
0 & 0 & f'(k)-2\gamma & 0 & 0 & -2\Omega\\
0 & 0 & \gamma_{L} & f'(k)-\gamma_{R} & 0 & 0\\
0 & 0 & 0 & 0 & g(k) & -\triangle\epsilon\\
0 & -\Omega & \Omega & 0 & \triangle\epsilon & g(k)
\end{array}\right)\; ,
\end{array}
\label{eq:Lmatrix}
\end{equation}
and the column
vector density matrix defined as
$\rho^{T}=(\rho_{00},\rho_{LL},\rho_{RR},\rho_{11},{\rm
  Re}\rho_{LR},{\rm Im}\rho_{LR})$.
 Here the matrix elements $\rho_{ab}=\rho_{ab}(k,q,t)$ with 
indices $a,b\in\{0,L,R,1\}$
denote the Fock states $\left|00\right\rangle$, $\left|10\right\rangle$, $\left|01\right\rangle$ and $\left|11\right\rangle $
of the system,i.e. no electron, one electron in the first dot (left dot),
one electron in the second dot (right dot), and one in each dot, respectively.
The functions 
$f(k)=D(e^{ik}-1)$, $f'(k)=D'(e^{ik}-1)$ and
$g(k)=\sqrt{DD'}e^{ik}-D_{a}-\Gamma$, where $D=\left|T\right|^{2}$ , $D'=\left|T+\chi\right|^{2}$, $\Gamma=({\gamma_{L}+\gamma_{R}})/{2}$,
$D_{0}=(D+D')/2$, $\triangle\epsilon=\epsilon_{2}-\epsilon_{1}$.

The factor 
\begin{equation}
\Gamma_{d}=\frac{(\sqrt{D'}-\sqrt{D})^{2}}{2}=\frac{\left|\chi\right|^{2}}{2}
\label{eq:dephasing_rate}
\end{equation} 
appears in
the diagonal elements of the last two rows of the resultant 
matrix $\mathcal{L}(k=0,q=0)$ of Eq.~(\ref{eq:Lmatrix})
and thus plays the role of
dephasing rate for the unconditional dynamics of the DQD's.
As $\Gamma_{d}$ becomes larger,
the QPC tends to localize the electron on the dot and thus 
reduces the coherent tunneling $\Omega$ that changes the DQD states
between $\left|01\right\rangle$ and $\left|10\right\rangle $.
When $\Omega\ll \Gamma_{d}$, one expects ${\rm Re}\rho_{LR} $
and ${\rm Im} \rho_{LR}$ from the last two rows of the master
equation, Eq.~(\ref{eq:LME}) with $\mathcal{L}(k=0,q=0)\rho$
defined in Eq.~(\ref{eq:Lmatrix}),
will decay much faster than other density matrix elements.
As a result, one can set the last two rows of Eq.~(\ref{eq:Lmatrix}))
equal to zero and then substitute the solution of
${\rm Re}\rho_{LR}$ and ${\rm Im}\rho_{LR}$
 back to the coupled equation.
Thus we obtain an effective tunneling rate between the two dots as
\begin{equation}
  \label{eq:Gamma_Omega}
\Gamma_{\Omega}=\frac{2\Omega^{2}/(\Gamma+\Gamma_{d})}{(1+(\frac{\Delta\epsilon}{\Gamma})^{2})}. 
\end{equation}
In this case, the $6\times6$ coherent tunneling matrix of
$\mathcal{L}(k,q)$ in the 
master equation in the Fourier space (counting field space)
reduces to a  $4\times4$ sequential tunneling matrix 
\begin{equation}
\begin{array}{l}
\mathcal{L}_{seq}(k,q)=\\
\left(\begin{array}{cccc}
f(k)-\gamma_{L} & 0 & \gamma_{R}e^{iq} & 0\\
\gamma_{L} & f(k)-\Gamma_{\Omega} & \Gamma_{\Omega} & \gamma_{R}e^{iq}\\
0 & \Gamma_{\Omega} & f'(k)-2\gamma-\Gamma_{\Omega} & 0\\
0 & 0 & \gamma_{L} & f'(k)-\gamma_{R}
\end{array}\right)
\end{array}
\label{eq:L_sequential}
\end{equation}
with $\Gamma_d$ defined in Eq.~(\ref{eq:dephasing_rate}), 
and the column
vector density matrix becomes
$\rho^{T}=(\rho_{00},\rho_{LL},\rho_{RR},\rho_{11})$
involved only the population elements.

In principle, one can solve the
resultant coupled first-order differential equations obtained from
Eq.~(\ref{eq:LME}) for the column elements of 
$\rho_{ab}(k,q,t)$ and then perform an inverse 
 Fourier transform to obtain $\rho_{ab}(N,M,t)$. The probability
 distribution of 
finding $N$ electron that have tunneled through the QPC and $M$ electrons
that have tunneled into the drain of the QD during time $t$
can then be obtained as:
\begin{eqnarray}
P(N,M,t) 
&=&{\rm Tr}_{\rm
  dot}[\rho(N,M,t)]
=\sum_{a}\rho_{aa}(N,M,t) \nonumber\\
&\hspace*{-1.5cm}=&
\hspace*{-0.9cm}\int_0^{2\pi}\int_0^{2\pi}\frac{dk
dq}{(2\pi)^2}e^{-ikN-iqM} \sum_a\rho_{aa}(k,q,t)
\label{PNMt}
\end{eqnarray}
From this distribution function $P(N,M,t)$, all orders of unconditional and 
conditional cumulants
(counting statistics) of transmitted electrons can be in principle calculated.

\section{Counting statistics: generating functional approach}
\label{sec:CS}

\subsection{Unconditional counting statistics}

In practice, a more efficient method is the  generating
functional technique. One may define the moment generating
function as \cite{Nazarov03}
\begin{equation}
e^{-F(k,q,t)}=\sum_{N.M} P(N,M,t)e^{ikN+iqM}.
\label{CGF}
\end{equation}
From this definition, it is easy to check that the $n$th moment of
$N$ and the $m$th moment of $M$ can be written as
\begin{equation}
  \label{eq:moment}
  \langle N^n M^m\rangle(t)=(-i\partial_k)^n(-i\partial_q)^m
  e^{-F(k,q,t)}|_{k=0=q}.
\end{equation}
The cross-cumulants can be calculated through the cumulant
generating function $F(k,q,t)$ as
\begin{equation}
  \label{eq:crosscumulant}
  \langle\langle N^n M^m\rangle\rangle(t)=- (-i\partial_k)^n
  (-i\partial_q)^m F(k,q,t)|_{k=0=q}.
\end{equation}
For example,  $\langle\langle O\rangle\rangle=\langle O \rangle$,
$\langle\langle O^2 \rangle\rangle=\langle (O-\langle O\rangle)^2 \rangle$,
$\langle\langle O^3 \rangle\rangle=\langle (O-\langle O\rangle)^3 \rangle$,
$\langle\langle O^4 \rangle\rangle=\langle (O-\langle O\rangle)^4 \rangle
-3\langle (O-\langle O\rangle)^2 \rangle$, etc.


From Eqs.~(\ref{CGF}), (\ref{PNMt}) and (\ref{rhokq}), the moment
generating function can then be obtained from $\rho(k,q,t)$ as:
\begin{equation}
e^{-F(k,q,t)}
= {\rm Tr}_{\rm dot}[\rho(k,q,t)]
=\sum_{a}\rho_{aa}(k,q,t),
\label{MGFkq}
\end{equation}
and the cumulant generating function is then
\begin{equation}
F(k,q,t)
=-\ln\left[\sum_a \rho_{aa}(k,q,t)\right].
\label{CGFkq}
\end{equation}
As a result, the unconditional
moments and cumulants can be calculated using Eqs.\ (\ref{MGFkq})
and(\ref{CGFkq}) according to Eqs.\ (\ref{eq:moment}) and
(\ref{eq:crosscumulant}). 

Thus the solution of the number-resolved master equation in the
Fourier space (counting field space) $\rho(k,q,t)$ has a direct
connection with the 
generating function approach to calculate the FCS. 

\subsection{Conditional counting statistics}

Having described the joint statistical properties of both the QPC
and QD currents, we discuss the conditional counting statistics: the
statistical current fluctuations (cumulants) of one system given the observation
of a given average current in the other system in time $t$.  

In QD-QPC transport system, the probability of having  $M$ electrons
tunneling into the drain of
the QD system conditioned on $N$ electrons passing through QPC in
time $t$ can be written as 
\begin{equation}
P(M|N,t)=P(N,M,t)/P(N,t)
\label{pcmnt}
\end{equation}
By defining the conditional moment generating function as 
\begin{equation}
e^{F_{c}(N,q,t)}\equiv\underset{M}{\sum}P(M|N,t)e^{iqM},
\label{mcgf}
\end{equation}
the $r$-th moment of electrons number $M$ passing through the QD system,
conditioned on the number of electrons $N$ in the drain of the QPC is given
by 
\begin{equation}
\left\langle M^{r}(t)
\right\rangle_{c}=\underset{M}{\sum}M^{r}P(M|N,t)
=\partial_{iq}^{r}e^{F_{c}(N,q,t)}|_{q=0},
\label{eq:mctD}
\end{equation}
where the subscript ``$c$'' denotes the quantity it attaches to being
conditional.
 The conditional cumulant $\left\langle \left\langle
     M^{r}(t)\right\rangle \right\rangle _{c}$ could be found by taking
partial derivatives with respect to $(iq)$ on the conditional
cumulant generating function $F_{c}(N,q,t)$.
\begin{equation}
  \label{eq:cumulant}
\left\langle \left\langle M^{r}(t)\right\rangle \right\rangle _{c} =  \partial_{iq}^{r}F_{c}(N,q,t)|_{q=0}  
\end{equation}
Using Eqs.~(\ref{pcmnt}) and (\ref{mcgf}), one observes that the
conditional cumulant generating function which is the logarithm
of the conditional moment generating function can be effectively
rewritten as 
\begin{equation}
F_{c}(N,q,t)=-\ln\ P(N,q,t),
\label{eq:Fc}
\end{equation}
where
\begin{eqnarray}
P(N,q,t)& = &{\sum_M}P(N,M,t)e^{iqM}\nonumber\\
&=&\frac{1}{2\pi}\int_{0}^{2\pi} dk P(k,q,t)e^{-ikN}. 
\label{eq:Pnqtc}
\end{eqnarray}
In obtaining Eq.~(\ref{eq:Pnqtc}), we have used the fact that 
$P(N,q,t)$ 
can also be expressed as the inverse Fourier transform
of $P(k,q,t)$ with respect to the counting field variable $k$.
Since $P(k,q,t)={\rm Tr}_{\rm dot}[\rho(k,q,t)]=\sum_a\rho_{aa}(k,q,t)$, 
one can calculate the conditional counting statistics once having  
the solution of the number-resolved master equation in the
Fourier space (counting field space) $\rho(k,q,t)$.

\subsection{FCS in the stationary state}

{\it Unconditional current cumulant.} 
In the stationary or steady state ($t\to\infty$), the calculation of
moments or cumulants can be simplified. 
The solution of Eq.\ (\ref{eq:LME}) can be symbolically written as
\begin{equation}
  \label{eq:soln}
  \rho(k,q,t)=e^{{\cal L}_{(k,q)}t}\rho(k,q,0).
\end{equation}
There is a unique eigenvalue $\lambda_1(k,q)$ of ${\cal L}(k,q)$ which
develops from the zero eigenvalue of ${\cal L}(k=0,q=0)$ with the smallest
absolute real part.
The rest of the eigenvalue(s) has (have) larger finite negative real
parts that 
make their contributions considerably much smaller for large times. As a
consequence,  the long-time dynamics of the moment generating
functional Eq.~(\ref{MGFkq}) near the stationary state can be well
approximated as \cite{Nazarov03,Sukhorukov07}
\begin{equation}
  \label{eq:ssMGFkq}
e^{-F(k,q,t)}
= {\rm Tr}_{\rm dot}[\rho(k,q,t)]
\approx e^{\lambda_1(k,q) t}.
\end{equation}
For the SQD-QPC system, the eigenvalue $\lambda_1(k,q)$ can be found
from Eq.\ (\ref{Lkq}) to be:
\begin{equation}
  \label{eq:egenvalue}
 \lambda_1(k,q)=(e^{ik}-1)D_0-\Gamma+\sqrt{[(e^{ik}-1)\Delta D-\Delta\Gamma]^2+\gamma_L\gamma_Re^{iq}},
\end{equation}
where $D_0=(D+D')/2$, $\Gamma=(\gamma_L+\gamma_R)/2$, $\Delta
D=(D-D')/2$ and $\Delta\Gamma=(\gamma_L-\gamma_R)/2$. Similarly, the
long-time (stationary-state) probability distribution
function from Eqs.~(\ref{PNMt}) and
(\ref{eq:ssMGFkq}) 
can be approximated as
\begin{equation}
 \label{eq:ssPNMtkq}
 P(N,M,t)
=\int_0^{2\pi}\int_0^{2\pi}\frac{dk dq}{(2\pi)^2}e^{-ikN-iqM+\lambda_1(k,q) t}.
\end{equation}
We can define the QPC current $I=N/t$ and QD current $J=M/t$ (setting $e=1$) in time $t$. Replacing $N=It$ and $M=Jt$, we then obtain the distribution function of the two current
\begin{equation}
 \label{eq:ssPIJtkq}
 P(I,J,t)
=\int_0^{2\pi}\int_0^{2\pi}\frac{dk dq}{(2\pi)^2}e^{[\lambda_1(k,q)-ikI-iqJ] t}.
\end{equation}
In the long-time (stationary) limit where the time $t$ should be much
larger than $\gamma^{-1}_{L,R}$, we may thus evaluate the integral
(\ref{eq:ssPIJtkq}) in the stationary phase approximation. The
dominant contribution to the joint probability distribution then takes
the form of a Legendre transform \cite{Sukhorukov07}:
\begin{equation}
  \label{eq:LT}
  \ln[P(I,J,t)]=t \min_{k,q}[\lambda_1(k,q)-ikI-iqJ].
\end{equation}

Since the long-time charge-number cumulant generating function
$F(k,q,t)$ from Eq.~(\ref{eq:ssMGFkq}) is linear in time,  we may define
the long-time (stationary-state) current cumulant generating function
as $\lambda_1(k,q)$, which is time-independent. The stationary-state current cumulant can then be calculated through
\begin{eqnarray}
  \label{eq:cumulantN}
\langle\langle I^n J^m\rangle\rangle
&=& \langle\langle N^n M^m\rangle\rangle/t \nonumber\\
&=&(-i\partial_k)^n
  (-i\partial_q)^m \lambda_1(k,q)|_{k=0=q}. 
\end{eqnarray}
Note that the time-dependence drops out in the expression of the
stationary-state current cumulant of Eq.~(\ref{eq:cumulantN}).

For example, the zero-frequency QD current noise and QPC current noise can also be calculated
\begin{eqnarray}
\langle \langle J^2 \rangle\rangle
&=&(-i\partial^2_q)\lambda_1(k,q)_{k=q=0}\nonumber\\
&=&\frac{\gamma _L \gamma _R \left(\gamma _L^2+\gamma
   _R^2\right)}{\left(\gamma _L+\gamma _R\right)^3},
\label{eq:QDnoise}\\
\langle \langle I^2 \rangle\rangle
&=&(-i\partial^2_k)\lambda_1(k,q)_{k=q=0} \nonumber\\
&=&\frac{2(D-D')^2 \gamma_L\gamma_R}{(\gamma_L+\gamma_R)^3}
\nonumber\\
&&+\frac{S(0)_{0}}{2}\left (\frac{\gamma_R}{\gamma_L+\gamma_R}\right )+\frac{S(0)_{1}}{2}\left (\frac{\gamma_L}{\gamma_L+\gamma_R}\right ),
\label{eq:QPCnoise}
\end{eqnarray}
where
$S(0)_0=2D$ and $S(0)_1=2D'$ are the values of the 
shot noise of the QPC for the QD (dot 2)
in $|0\rangle$ and $|1\rangle$ states, respectively. 
The first term of the QPC current noise of Eq.~(\ref{eq:QPCnoise})
comes from the 
random telegraph process in the QPC currents making transitions between
$D$ and $D'$ caused by the electrons randomly
tunneling onto and out of dot 2 with rates $\gamma_L$ and
$\gamma_R$, respectively.

{\it Conditional current cumulant.}
Similarly, in the stationary state, the conditional current cumulant
generating function $\lambda_{cI}(I,q)$ and $\lambda_{cJ}(k,J)$ can be calculated
from the reverse partial Fourier transform of the
joint generating function\cite{Sukhorukov07}
\begin{eqnarray}
 \label{eq:ssPIqkJt}
e^{ \lambda_{cI}(I,q) t}=P(I,q,t)
&=&\int_0^{2\pi}\frac{dk}{(2\pi)}e^{[\lambda_1(k,q)-ikI] t}, \\
e^{ \lambda{cJ}1(k,J) t}=P(k,J,t)
&=&\int_0^{2\pi}\frac{dq}{(2\pi)}e^{[\lambda_1(k,q)-iqJ] t}.
\end{eqnarray}
One may evaluate the integral in the stationary phase approximation to
obtain\cite{Sukhorukov07}
\begin{eqnarray}
  \label{eq:conditionalCGF}
  \lambda_{cI}(I,q)&=&\min_{k}[\lambda_1(k,q)-ikI], 
\label{lambda_cI}\\
  \lambda_{cJ}(k,J)&=&\min_{q}[\lambda_1(k,q)-iqJ].
\label{lambda_cJ}
\end{eqnarray}
The conditional current cumulant then can be calculated from
\begin{eqnarray}
\langle\langle J^m\rangle\rangle_c=
  (-i\partial_q)^m \lambda_{cI}(I,q)|_{q=0}.
 \label{eq:conditionalCCJ} \\
 \langle\langle I^n\rangle\rangle_c=
  (-i\partial_k)^n \lambda_{cJ}(k,J)|_{k=0}.
 \label{eq:conditionalCCI}
\end{eqnarray}

However, the conditional current cumulant
generating functions of Eqs.~(\ref{lambda_cI}) and (\ref{lambda_cJ})
are difficult to evaluate unless an analytic form of the eigenvalue
$\lambda_1(k,q)$ is available.
Even for the problem of the SQD-QPC system where the eigenvalue 
$\lambda_1(k,q)$ can be obtained analytically, it is still not easy to
evaluate Eqs.~(\ref{lambda_cI}) and (\ref{lambda_cJ}) directly. 
In Ref.~\onlinecite{Sukhorukov07}, a further approximation to neglect the shot
noise contribution from the QPC was made. As a result, the calculations are
significantly simplified and the conditional generating
functions and conditional current cumulants were obtained in analytical
forms. 
From the zero-frequency noise of Eq.~(\ref{eq:QPCnoise}),
the QPC shot noise terms (the last two terms) can be neglected as
compared to the first term when
the parameters $(D-D')^2\gg (D,D')(\gamma_{L}+\gamma_{R})$. In
Ref.~\onlinecite{Sukhorukov07}, the QPC tunneling rates are set to be
$D'\approx5.03\times 10^{10}$ Hz and $D\approx 4.85\times 10^{10}$ Hz, 
and the QD tunneling rates are chosen as $\gamma_L=160$ Hz, $\gamma_R=586$ Hz in
configuration A and $\gamma_L=512$ Hz, $\gamma_R=345$ Hz in
configuration B. 
So it was valid to neglect the QPC shot noise terms for the parameters
used in Ref.~\onlinecite{Sukhorukov07}. 
Neglecting the QPC shot noise terms amounts to replacing
$e^{ik}\rightarrow1+ik$ in Eq.~(\ref{eq:egenvalue}).  
Consequently, the conditional current cumulant
generating functions of Eqs.~(\ref{lambda_cI}) and
(\ref{lambda_cJ}) can be obtained analytically, so are the
conditional current cumulants.
However, when the QPC shot noise terms cannot be neglected, 
the conditional steady-state generating functions of Eqs.~(\ref{lambda_cI}) and
(\ref{lambda_cJ}) and thus also the conditional current 
cumulants (\ref{eq:conditionalCCJ}) and (\ref{eq:conditionalCCI})
are difficult to
obtain even using 
the analytic form of the eigenvalue of
Eq.~(\ref{eq:egenvalue})
due to the fact that the numerical minimization and then
numerical derivatives 
that need to be performed are quite numerically unstable. 
It is even more difficult
for more complicated interacting
nanoscale conductors with the dimension of the matrix
equation of the master equation growing quickly 
and no analytical forms of
eigenvalues of ${\cal L}(k,q)$ are available. 

\subsection{Efficient numerical method} 
\label{sec:EffNumMeth} 
It is thus desirable to develop an efficient and numerically stable 
method to calculate the
conditional counting statistics for a wider range of parameters and  
for more complicated interacting quantum transport systems.  
For unconditional steady-state cumulants,  the projection operator
technique with perturbation expansion in counting fields
developed in Refs.~\onlinecite{key-6} and \onlinecite{Flindt04,Flindt05,Flindt08,Lambert13}
can be used to circumvent the problems of large system dimensions and 
the instability of taking 
numerical derivatives on the generating function.
However, things are different in the conditional case.
The unconditional cumulants are evaluated in the counting field
(inverse Fourier transform) space, e.g.,
the steady-state cross current cumulant of Eq.~(\ref{eq:cumulantN}).
Thus 
a perturbation partition of
the Liouvillian matrix ${\cal L}(k,q)$ in Eq.~(\ref{eq:LME})
can be performed to calculate corrections
to the maximum eigenvalue [with counting fields set to zero, 
e.g., $\lambda_1(k=0,q=0)$] 
 order-by-order in the counting fields
to avoid taking derivatives.
In contrast, the conditional cumulants are evaluated in the 
partial or mixed Fourier transform
space, i.e., the mixed space of counting field of one system and tunneled electron number of the other system, e.g., the conditional current cumulant of
Eq.~(\ref{eq:conditionalCCJ}) in which $I=N/t$. 
Thus even though the perturbation expansion can be performed in the
counting field $q$, the Liouvillian matrix ${\cal L}(N,q)$ in the
mixed $N$-resolved and counting field master equation will couple 
the $N$-sector density matrix elements with the $(N-1)$-sector ones,
forming a huge coupled difference equations that are difficult to solve.
To proceed, 
one crucial observation is that the conditional moment
of Eq.~(\ref{eq:mctD}), with the help of Eqs.~(\ref{pcmnt}) , (\ref{mcgf})  and (\ref{eq:Pnqtc}),
can be written as 
\begin{equation}
\left\langle M^{r}(t)\right\rangle_{c}
=\frac{1}{2\pi P(N,t)}\int_{0}^{2\pi} dk\,  {\rm Tr}_{\rm
  dot}[\partial_{iq}^{r}\rho(k,q,t)|_{q=0}] e^{-ikN},
\label{eq:mct}
\end{equation}
where $P(N,t)=\frac{1}{2\pi}\int_{0}^{2\pi} dk\, 
{\rm Tr}_{\rm  dot}[\rho(k,q,t)|_{q=0}] e^{-ikN}$.
Thus if we can find out how  the $r$th
derivatives
$\partial_{iq}^{r}\rho(k,q,t)|_{q=0}$ evolve in time $t$ directly,
then we can 
just perform the 
trace and inverse Fourier transform to obtain directly the
moments of electron number through the QD system conditioned on a
given QPC current. 

To find the evolution equations for the variables 
$\partial_{iq}^{r}\rho(k,q,t)|_{q=0}$,  
let us take partial derivatives with respect to the counting factor
$iq$ on Eq.~(\ref{eq:LME}) $r$ times for the QD system and then set $q=0$. 
We obtain $r$ differential equations as \cite{key-31}
\begin{equation}
\begin{array}{c}
\dot{\rho}(k,q,t)|_{q=0}=\mathcal{L}(k,q)\rho(k,q,t)|_{q=0}\\
\partial_{iq}\dot{\rho}(k,q,t)|_{q=0}=\partial_{iq}[\mathcal{L}(k,q)\rho(k,q,t)]|_{q=0}\\
\partial_{iq}^{2}\dot{\rho}(k,q,t)|_{q=0}=\partial_{iq}^{2}[\mathcal{L}(k,q)\rho(k,q,t)]|_{q=0}\\
\vdots\\
\partial_{iq}^{r}\dot{\rho}(k,q,t)|_{q=0}=\partial_{iq}^{r}[\mathcal{L}(k,q)\rho(k,q,t)]|_{q=0}\;.
\end{array}
\label{eq:DerMatrix}
\end{equation}
Note that 
with the expression of $\mathcal{L}(k,q)$ available [ e.g., given by Eq.~(\ref{Lkq})
or (\ref{eq:Lmatrix})], the derivatives of 
\begin{eqnarray}
\partial_{iq}^{r}[\mathcal{L}(k,q)\rho(k,q,t)]|_{q=0} 
&=&\{[\partial_{iq}^{r}\mathcal{L}(k,q)]\rho(k,q,t)\}|_{q=0}\nonumber\\
&&\hspace{-1cm}+\{\mathcal{L}(k,q)[\partial_{iq}^{r}\rho(k,q,t)]\}|_{q=0} 
\end{eqnarray}
 in Eq.~(\ref{eq:DerMatrix}) should be evaluated first.  
Then the equation resulting from Eq.~(\ref{eq:DerMatrix}) forms a set
of coupled differential 
equations for variables  
$\partial_{iq}^{r}\rho(k,q,t)|_{q=0}$.
In this way, the derivatives of 
$\partial_{iq}^{r}\rho(k,q,t)|_{q=0}$ can be considered as being 
performed beforehand as one can obtain the solutions for the $r$th
derivatives  $\partial_{iq}^{r}\rho(k,q,t)|_{q=0}$ directly \cite{key-31}
and can thus avoid taking the derivatives later on
the generating functions (if the generating functions were obtained numerically first), which
are often quite numerically unstable. 
We can define a super-vector $\sigma$ as 
\begin{equation}
  \label{eq:sigma}
 \sigma(k,q=0,t)=\left(
\begin{array}{c}
\rho(k,q,t)|_{q=0}\\
\partial_{iq}\rho(k,q,t)|_{q=0}\\
\partial_{iq}^{2}\rho(k,q,t)|_{q=0}\\
\vdots\\
\partial_{iq}^{r}\rho(k,q,t)|_{q=0}
\end{array}\right)\; . 
\end{equation}
and write Eq.~(\ref{eq:DerMatrix}) as 
\begin{equation}
\dot{\sigma}(k,q=0,t)=Z(k,q=0)\sigma(k,q=0,t)
\label{eq:EOMsigma}
\end{equation}
where the $Z(k,q=0)$ matrix contains all the elements of $\mathcal{L}$ and
its partial derivatives. 
The solution of Eq.~(\ref{eq:EOMsigma}) can be obtained by the
exponentiation of the $Z(k,q=0)t$ matrix as
\begin{equation}
\sigma(k,q=0,t)=e^{Z(k,q=0) t}\sigma(k,q=0,t=0).
\end{equation}
Performing the inverse
Fourier transform of the super-vector $\sigma(k,q=0,t)$,
then tracing over the
system degree of freedom on the $r$-th derivative components of the
resultant super-vector and finally divide the quantity by the
probability $P(N,t)$, one obtains the conditional moment 
$\left\langle  M^{r}\right\rangle_{c}$. 
Note that $P(N,t)$ is just the trace of the $0$-th derivative components of the
resultant inverse-Fourier-transformed super-vector over the system
degrees of freedom. 
The conditional current cumulants can be obtained from the conditional
moments $\left\langle  M^{r}\right\rangle_{c}$. 
For example, 
the first and second conditional current cumulants are obtained by 
\begin{eqnarray}
\left\langle \left\langle J\right\rangle \right\rangle_{c}
&=&\frac{\left\langle M\right\rangle _{c}}{t}
\label{eq:Jc} \\
\left\langle \left\langle J^{2}\right\rangle \right\rangle _{c}
&=&\frac{\left\langle M^{2}\right\rangle_c -\left\langle
    M\right\rangle_c ^{2}}{t}\;. 
\label{eq:J2c} 
\end{eqnarray}
This method can be applied to deal with more complex systems with
larger dimension of the $Z(k,q)$ matrices.

In summary, to avoid numerical instability and complexity of taking
derivatives on the generating functions, we develop an efficient
numerical method to calculate the conditional moments and
cumulants for more complicated interacting quantum transport systems. 
This method also allows the calculations of both transient and
stationary conditional counting statistics.
To demonstrate its advantage and usage, we will apply this method to
calculate the first and second current cumulants of two nanoscale
interacting conductor systems.
The first one is just the SQD-QPC system but 
without ignoring the QPC shot noise.
The second one is a more complicated system of
DQD's in series with one of the dots measured by a QPC, for which
analytical eigenvalues of matrix $\mathcal{L}(k,q)$ of the evolution
equation for general parameters are not available.

Considering, for example, the QPC shot noise without replacing
$e^{ik}\rightarrow1+ik$ for the SQD-QPC system, we can write 
in our numerical method the matrix form 
$\dot{\sigma}(k,q,t)=\mathcal{Z}(k,q)\sigma(k,q,t)$
to calculate the first and second QD current moments and cumulants 
 conditioned on the QPC current $I$,
where 
\begin{equation}
\sigma(k,q,t)=\left(\begin{array}{c}
\rho_{00}(k,q,t)\\
\rho_{11}(k,q,t)\\
\partial_{iq}\rho_{00}(k,q,t)\\
\partial_{iq}\rho_{11}(k,q,t)\\
\partial_{iq}^{2}\rho_{00}(k,q,t)\\
\partial_{iq}^{2}\rho_{11}(k,q,t)
\end{array}\right)
\end{equation}
 and 
\begin{equation}
\begin{array}{l}
\mathcal{Z}(k,q)=\\
\left(\begin{array}{cccccc}
f_{1}(k) & \gamma_{R}e^{iq} & 0 & 0 & 0 & 0\\
\gamma_{L} & f_{2}(k) & 0 & 0 & 0 & 0\\
0 & \gamma_{R}e^{iq} & f_{1}(k) & \gamma_{R}e^{iq} & 0 & 0\\
0 & 0 & \gamma_{L} & f_{2}(k) & 0 & 0\\
0 & \gamma_{R}e^{iq} & 0 & 2\gamma_{R}e^{iq} & f_{1}(k) & \gamma_{R}e^{iq}\\
0 & 0 & 0 & 0 & \gamma_{L} & f_{2}(k)
\end{array}\right)\; ,
\end{array}
\end{equation}
with $f_{1}(k)=D(e^{ik}-1)-\gamma_{L}$ and
$f_{2}(k)=D'(e^{ik}-1)-\gamma_{R}$.

\section{Result and discussion}
\label{sec:Results}
We focus on the first two orders of the QD current cumulants
i.e., $\left\langle \left\langle J\right\rangle \right\rangle _{c}$
and $\left\langle \left\langle J^{2}\right\rangle \right\rangle _{c}$,
conditioned on an observed QPC current $I$.   
We will vary the QPC tunneling rates such that the difference in the
QPC tunneling rates with and without the occupation of an electron on
dot 2 goes from high to low, and at the same time the
condition to ignore the QPC shot noise term is also progressively not
satisfied. We will take the QD tunneling rates to be $\gamma_L=160$ Hz,
$\gamma_R=586$ Hz which are the same as those of configuration A in
Ref.~\onlinecite{Sukhorukov07}.    
As stated earlier,  the QPC shot noise terms of the second
and third terms of Eq.~(\ref{eq:QPCnoise}) can be neglected as
compared to the first term of the random-telegraph-process noise when 
$(D'-D)^2\gg (D,D')(\gamma_L+\gamma_R)$.
Since the values of 
$(D',D)=(5.03\times10^{10},\:4.85\times10^{10})$ Hz 
have already been
demonstrated in Ref.~\onlinecite{Sukhorukov07} to be an excellent
parameter set  
to neglect the QPC shot noise,
we specifically choose three sets of the 
QPC tunneling rates to be $(D',D)=(5.03\times10^{8},\:4.85\times10^{8})$ Hz,
$(5.03\times10^{7},4.85\times10^{7})$ Hz,
$(5.03\times10^{6},\:4.85\times10^{6})$ Hz to investigate the effect of shot
noise.
Note that these values of $(D,D')$ in front of the exponents in the
three sets of $(D,D')$ are the
same. As a result, when the 
QPC tunneling rate (i.e., the exponent) decreases, the random
telegraph signal 
will no longer dominate over the QPC shot noise contribution and therefore the
QPC shot noise cannot be neglected completely.
In other words, the analytic method that replaces
$e^{ik}\rightarrow 1+ik$ in Ref. \onlinecite{Sukhorukov07} 
will be progressively not valid as the 
QPC tunneling rates in the
three sets of $(D,D')$ decrease from high to low. 

\subsection{SQD-QPC system}
We will show how the QPC shot noise (intrinsic
noise) affects the conditional current cumulants for the SQD-QPC system
through the quantities of the joint probability distribution
$P(I,J,t)$ of
detecting the QPC current $I$ and QD current $J$, 
 and the conditional current
($\left\langle \left\langle J\right\rangle \right\rangle _{c}$ and
conditional zero-frequency noise
 $\left\langle \left\langle J^{2}\right\rangle \right\rangle _{c}$.

\begin{figure}
\includegraphics[width=8cm]{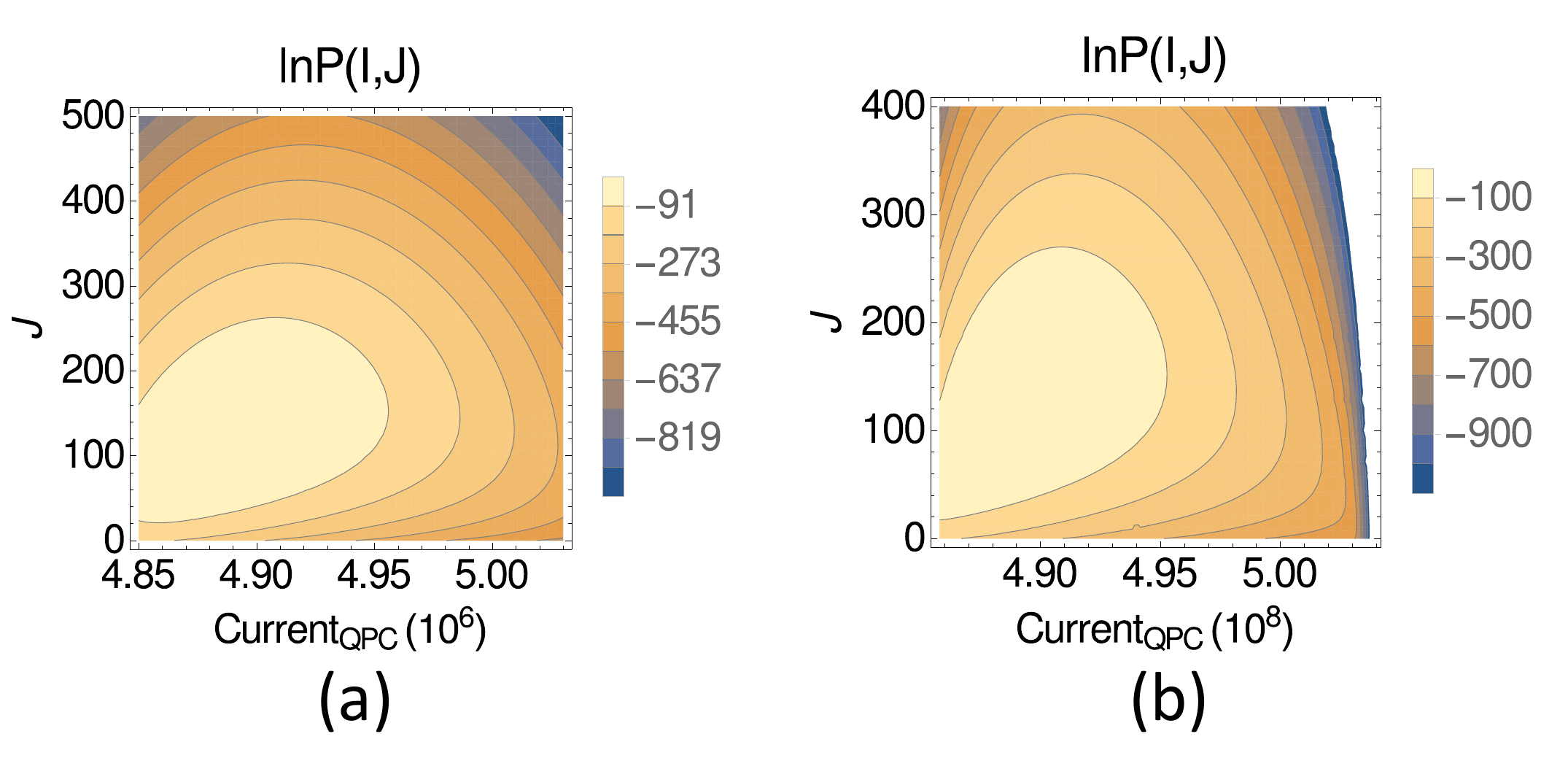}
\caption{The logarithm of the joint probability distribution,
  $\ln P(I,J)$, of 
observing SQD current $J$ and QPC current $I$ as a color contour plot
obtained by using maximum eigenvalue
method for (a)  $(D',D)=(5.03\times10^{6},4.85\times10^{6})$ Hz 
(b) $(D',D)=(5.03\times10^{8},4.85\times10^{8})$ Hz. 
The tunneling rates of the SQD are $\gamma_{L}=160$ Hz and
$\gamma_{R}=586$ Hz. }
\label{fig:PIJT}
\end{figure}

\subsubsection{Joint probability distribution}
\label{sec:jointProb}

With the maximum eigenvalue  $\lambda_1(k,q)$ 
of Eq.~(\ref{eq:egenvalue}) available, it is possible to obtain the joint
probability distribution $P(I,J)$ from Eq.~(\ref{eq:LT}) 
without replacing $e^{ik}\rightarrow1+ik$.
The contour plots of the logarithm of the joint
probability distribution $\ln P(I,J)$ 
of detecting QPC current $I$ (horizontal axis) and
QD current $J$ (vertical axis) for $(D',D)=(5.03\times10^{6},4.85\times10^{6})$ Hz and 
 $(D',D)=(5.03\times10^{8},4.85\times10^{8})$ Hz are shown in 
Figs.~\ref{fig:PIJT}(a) and \ref{fig:PIJT}(b), respectively.
The logarithm of the joint
probability distribution in 
Fig.~\ref{fig:PIJT}(a), with low QPC tunneling rates $(D,D')$ 
for which $(D'-D)^2$ is comparable to $ (D,D')(\gamma_L+\gamma_R)$,
differs considerably from that of neglecting the QPC
current shot noise presented in Ref.~\onlinecite{Sukhorukov07}.
Especially near the endpoints of $I=D$ and $I=D'$, substantially larger
probabilities for finite $J$ values are observed here.
This indicates (see the discussion in the next paragraph about
the conditional current probability)
that the resultant conditional QD current and noise 
conditioned on the observed QPC current at or near $I=D$ and
$I=D'$ will deviate
from zeros as those shown in Ref.~\onlinecite{Sukhorukov07}. 
For the parameter set of higher QPC tunneling rates
shown in Fig.~\ref{fig:PIJT}(b), the  the logarithm of joint current
probability distribution 
looks closer to that of Ref.~\onlinecite{Sukhorukov07}.

The conditional QD 
current probability $P(J|I,t)$ can be obtained from the joint
current probability distribution $P(I,J,t)$
using the Bayesian formalism as
\begin{equation}
P(J|I,t)=\frac{P(I,J,t)}{P(I,t)}=\frac{P(I,J,t)}{\int dJ\, P(I,J,t)}
\label{eq:PJ|I}
\end{equation}
With conditional current probability, we can calculate the conditional
quantities. 
The conditional current moments  can
  be obtained directly from $P(J|I,t)$ as $\langle
  J^{r}(t)\rangle _{c}\equiv\intop_{0}^{\infty}dJ\, P(J|I,t)J^{r}(t)$
and
  the conditional current cumulants $\langle\langle
  J^{r}(t)\rangle \rangle_{c}$ can be calculated
   from  $\langle  J^{r}(t)\rangle_{c}$ accordingly, e.g., 
$\left\langle \left\langle J^{2}\right\rangle \right\rangle _{c}=\left\langle J^{2}\right\rangle _{c}-\left\langle J\right\rangle^2_{c}$. 
We will show in the next section the results of the conditional QD
current and zero-frequency  
noise (the first and second conditional cumulants)
using the method of the joint current probability distribution 
and the Bayesian formalism as a confirmation of our
numerical method for the SQD-QPC case.

Integrating the conditional $\left\langle J^{r}(t)\right\rangle_{c}$
over the QPC current probability $P(I,t)$ gives the corresponding
unconditional current moments: $\langle J^r(t)\rangle
=\intop_{0}^{\infty}dI\, P(I,t)\langle J^{r}(t)\rangle_{c}$.
This formula demonstrates that the conditional quantities,
$\left\langle J^{r}(t)\right\rangle_{c}$, provides us
with more information and can give
insight into the unconditional quantities.


\subsubsection{First and second conditional current cumulants}

\begin{figure}
\includegraphics[width=8cm]{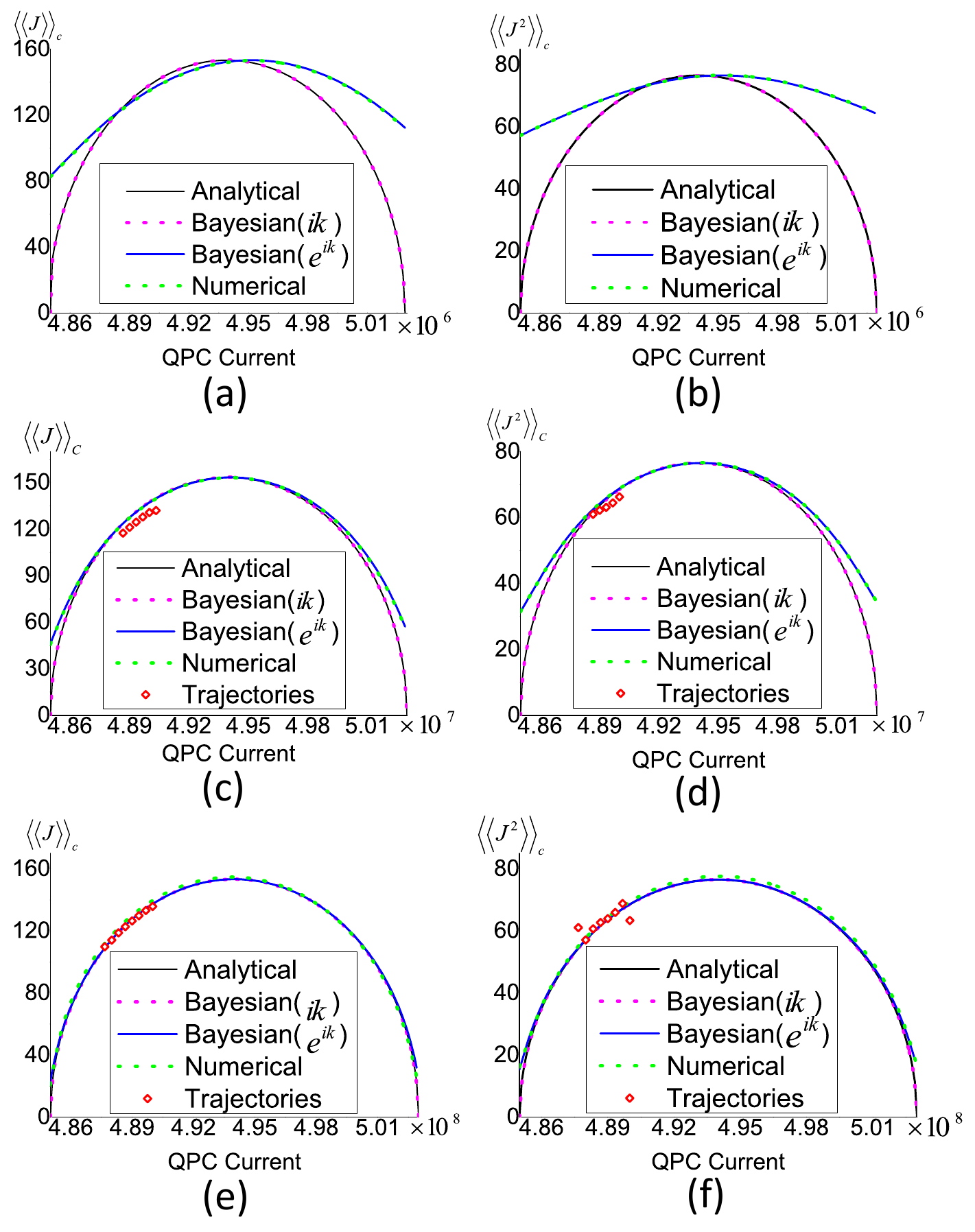}

\caption{Conditional SQD current, the first cumulant, (left panel)
  and zero-frequency noise, the second cumulant, (right panel) obtained by different methods for  
(a) and (b) $(D',D)=(5.03\times10^{6},4.85\times10^{6})$ Hz,
(c) and (d) $(D',D)=(5.03\times10^{7},4.85\times10^{7})$ Hz, and
(e) and (f) $(D',D)=(5.03\times10^{8},4.85\times10^{8})$ Hz.
The tunneling rates of the SQD are $(\gamma_{L},\gamma_{R})=(160,586)$ Hz.
}
\label{fig:SQD_cumulants}
\end{figure}

The first and second steady-state conditional QD current cumulants $\left\langle \left\langle J\right\rangle \right\rangle _{c}$
and $\left\langle \left\langle J^{2}\right\rangle \right\rangle _{c}$ shown
in Figs.~\ref{fig:SQD_cumulants}(a) and~\ref{fig:SQD_cumulants}(b) for low
QPC tunneling rates $(D',D)=(5.03\times10^{6},4.85\times10^{6})$ Hz,
Figs.~\ref{fig:SQD_cumulants}(c) and \ref{fig:SQD_cumulants}(d) for
medium tunneling rates $(D',D)=(5.03\times10^{7},4.85\times10^{7})$ Hz and
Fig.~\ref{fig:SQD_cumulants}(e) and \ref{fig:SQD_cumulants}(f) for
high tunneling rates
$(D',D)=(5.03\times10^{8},4.85\times10^{8})$ Hz
are obtained by five different methods:
(i) the analytical formulas neglecting the QPC shot noise given in
Ref.~\onlinecite{Sukhorukov07} (in thin solid line),  
(ii) the joint probability $P(I,J,t)$ obtained
with the replacement of $(e^{ik}\rightarrow1+ik)$ (i.e.,
neglecting the QPC shot noise) and the 
Bayesian rules (in light-blue dotted line), 
(iii) the joint probability $P(I,J,t)$ obtained without the
approximation of $(e^{ik}\rightarrow1+ik)$ and the 
Bayesian rules (in red open triangles), 
(iv) the numerical method described in Sec.~\ref{sec:EffNumMeth} (in
blue dots) and   
(v) the quantum trajectory method described in Sec.~\ref{sec:Qtrajectory}
(in red open diamonds).
As expected, the curves obtained by methods (i) and (ii) coincide and by
methods (iii) and (iv) coincide for a given set of QPC tunneling rates.    
They all approach to each other for the  
high QPC tunneling rate case in    
Figs.~\ref{fig:SQD_cumulants}(e) and \ref{fig:SQD_cumulants}(f), then start to deviate from each other near 
 the endpoints of $I=D$ and $I=D'$ for the medium  QPC tunneling rate
 case  in Figs.~\ref{fig:SQD_cumulants}(c) and
 \ref{fig:SQD_cumulants}(d) and differ significantly from each other for the low QPC tunneling rate
 case  in Figs.~\ref{fig:SQD_cumulants}(a) and \ref{fig:SQD_cumulants}(b). 
This is consistent with
 the observations of the joint current probability distribution discussed    
in Sec.~\ref{sec:jointProb} and indicates that one can approximately 
neglect the QPC
shot noise for the parameter set shown in
Figs.~\ref{fig:SQD_cumulants}(e) and \ref{fig:SQD_cumulants}(f)
but cannot do so for the parameter set shown in
Figs.~\ref{fig:SQD_cumulants}(a) and \ref{fig:SQD_cumulants}(b).
When the QPC shot noise is completely ignored,
$\left\langle \left\langle J\right\rangle \right\rangle _{c}$ and
$\left\langle \left\langle J^{2}\right\rangle \right\rangle _{c}$
calculated by analytical solution are universal semicircles as a function
of the current $I$, have a maximum 
at $I=({D+D'})/{2}$ 
 and equal to zero at $I=D,D'$ 
even though the SQD tunneling rates are asymmetric, i.e.,
$\gamma_L\neq\gamma_R$ \cite{Sukhorukov07}. 
But when the QPC shot noise is not negligible, 
when QPC current $I=D,D'$, the SQD is no longer completely occupied or
empty within the whole duration time $t$. 
As a result, the conditional SQD 
$\left\langle \left\langle J\right\rangle \right\rangle _{c}$ and
$\left\langle \left\langle J^{2}\right\rangle \right\rangle _{c}$
are not equal to zero at the endpoints $I=D,D'$ of the interval
$I=[D,D']$.
Furthermore $\left\langle \left\langle J\right\rangle \right\rangle _{c}$
and $\left\langle \left\langle J^{2}\right\rangle \right\rangle _{c}$
in this case become asymmetric for different $\gamma_L$ and
$\gamma_R$ tunneling rates.
At the QPC current $I=D$, the SQD in most of the
duration time $t$ is empty, 
while at the QPC current $I=D'$, the SQD in most of the duration time
$t$ is occupied.
For the parameter $(\gamma_{L},\gamma_{R})=(160,586)$ of Fig.~\ref{fig:SQD_cumulants},
the SQD has a larger unconditional probability
${\gamma_{R}}/({\gamma_{L}+\gamma_{R}})$ of
being empty than the probability
$\gamma_{L}/({\gamma_{L}+\gamma_{R}})$ of
being occupied. 
This leads to more switchings at $I=D'$ than at $I=D$
and thus results in larger 
conditional current cumulants
 $\left\langle \left\langle J\right\rangle \right\rangle _{c}$
and $\left\langle \left\langle J^{2}\right\rangle \right\rangle _{c}$
at $I=D'$ than at $I=D$ 
with maximums occurring  at  $I>({D+D'})/{2}$.
The results becomes opposite if $\gamma_L>\gamma_R$  for which  
the SQD has a larger unconditional probability
${\gamma_{L}}/({\gamma_{L}+\gamma_{R}})$ of
being occupied, leading to more switchings at $I=D$ than at $I=D'$.
Thus the conditional current cumulant $\left\langle \left\langle
    J\right\rangle \right\rangle _{c}$ 
and $\left\langle \left\langle J^{2}\right\rangle \right\rangle _{c}$
at $I=D$ are larger than at $I=D'$ 
with maximums occurring at $I<({D+D'})/{2}$.
If $\gamma_L=\gamma_R$, the
equal unconditional 
probability of being empty and being occupied makes the QPC shot
noise contribution symmetric with respect to $I=({D+D'})/{2}$, 
resulting in symmetrical
conditional current cumulants   
$\left\langle \left\langle J\right\rangle \right\rangle _{c}$
and $\left\langle \left\langle
    J^{2}\right\rangle\right\rangle _{c}$ with maximums at $I=({D+D'})/{2}$.

We also simulate 120,000 realizations of the conditional SQD occupation
number $\langle n_2(t) \rangle_c$
and QPC currents by quantum trajectory method  to calculate 
$\langle\langle J\rangle\rangle _{c}$ 
and $\langle\langle J^2 \rangle \rangle _{c}$. The quantum trajectory
method is described in the next section.
     
\subsection{Counting statistics by quantum trajectories}
\label{sec:CCSbyQT}

\begin{figure}
\includegraphics[width=9cm]{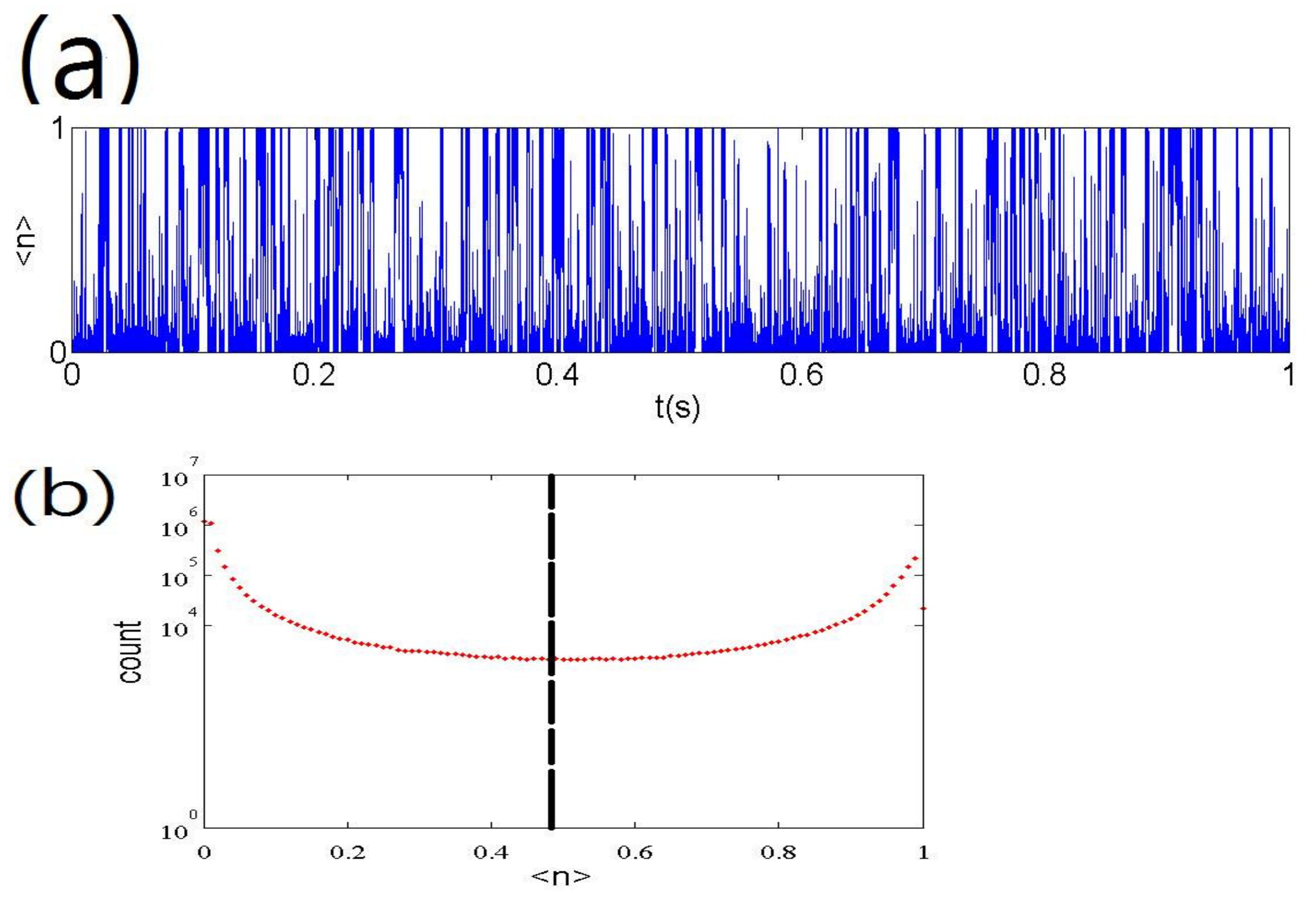}
\caption{(a) A simulated 
trajectory of $\left\langle n_2\right\rangle_c $ for relatively moderate
QPC tunneling rates of
$(D,D',\gamma_{L},\gamma_{R})=(4.85\times10^{7},5.03\times10^{7},160,586)$ Hz, i.e., with a moderate induced dephasing rate.
(b) Histogram of each interval value of conditional $\left\langle n_2\right\rangle_c $
for the trajectory shown in (a). 
The values form $0$ to $1$ for $\left\langle n_2\right\rangle_c $ of the
horizontal axis is
divided into 101 intervals and the vertical axis of the number of
counts is plotted on a logarithmic scale.  
The dashed line in the middle is
the reference line of $\left\langle n_2\right\rangle_c $.
}
\label{fig:histogram}
\end{figure}

We describe how we obtain the conditional
counting statistics using the quantum trajectory method. Take the
case of SQD-QPC as an example. 
Using Eqs.~(\ref{dn_c}) and (\ref{I_c}), we can numerically simulate
the evolutions of 
the conditional expectation value of the electron occupation number 
$\langle  n_2\rangle_{c}(t)$ on the QD as well as the measured conditional
instantaneous QPC current record $I_c(t)$  
 in a single run of a realistic experiment as shown in 
Fig.~\ref{fig:Fig2}(b).
The time average QPC current $I$ in
time $t$ can be obtained by integrating the instantaneous QPC current
$I_c(t)$ over time $t$.
The average QD current $J$ can be obtained by the number  $M$
of electrons
 transmitted through the QD in time $t$, i.e., $J=M/t$.    
The number $M$ could be determined \cite{Sukhorukov07} by the number of ``up'' and then
immediate ``down'' switches $M$ of the random telegraph signal 
in a given time trace of $I_c(t)$ of
duration $t$,
or the number of ``1'' and then
immediate ``0'' switches $M$ in a given time trace 
$\langle n_2 \rangle_c(t)$  
of duration $t$ [see Fig.~\ref{fig:Fig2}(a)].    
When the strength of the random telegraph signal is much larger than
that of the QPC intrinsic 
current (shot) noise \cite{Sukhorukov07}, this provides an excellent
way to determine the occupation number $\langle n_2 \rangle_c(t)$ on the
QD and the average current $J$ through the QD (even if the current $J$ is rather weak).
However, when the noise induced by the random telegraph signal is not
much smaller than or 
is comparable to the QPC shot noise, the measured QPC current  $I_c(t)$ 
may not be able to give an unambiguous  
measurement of the occupation number '1' or '0' on the QD.
For example, a typical trace or realization of $\langle n_2 \rangle_c(t)$ 
for the parameter set of medium QPC tunneling rates of 
$(D',D)=(5.03\times10^{7},4.85\times10^{7})$ Hz is shown in 
Fig.~\ref{fig:histogram}(a). 
Comparing the QPC current trace shown in Fig.~\ref{fig:histogram}(a)
to that in Fig.~\ref{fig:Fig2}(a) for
the case of 
$(D',D)=(5.03\times10^{10},4.85\times10^{10})$ Hz with a large $\chi$
value,
one finds that 
many values of $\langle n_2 \rangle_c(t)$ in the
trajectory shown in
Fig.~\ref{fig:histogram}(a) are close to or in-between but
 not at either $0$ or $1$ 
due to relatively large contribution from the shot noise fluctuations.  
This leads to ambiguity in
counting the number of electrons tunneling on and off dot 2 and then onto the
right lead (or drain) with the method outlined in
Ref.~\onlinecite{Sukhorukov07}.    
Therefore, we adopt
a semiempirical method with details described in Appendix
\ref{App:counting_method} to
diminish the ambiguity, 
to count the number  $M$ 
of electrons which have tunneled through dot 2 into the right
lead of the QD system, and thus to obtain the average 
QD current $J={M}/{t}$
for a given trajectory 
 $\langle n_2 \rangle_c(t)$ in time $t$. 
Moreover, the corresponding average QPC current $I$ in
time $t$ can be obtained by integrating the instantaneous QPC current
of $I_c(t)$ generated from Eq.~(\ref{I_c}).
Thus we know both $J$ and $I$ of the given trajectory 
$\langle n_2 \rangle_c(t)$ in time $t$. 
We divide the values of $I(J)$ into small intervals.  
After simulating many trajectories to get meaningful statistics,
we categorize the values of $I(J)$ extracted from the trajectories
into their corresponding small intervals. 
To calculate the QD current cumulant $\langle \langle J \rangle
\rangle_{c}$ conditioned on the QPC current $I$, we calculate the
average of the different values of the QD currents $J_i$ 
 that all fall into the same small interval  $I_i$ (i.e.,
corresponding to the
same value of $I$) as $\langle \langle J \rangle
\rangle_{c}= \sum_{i=1}^{S}J_{i}/S$, where $S$ is the number of
samples $J_i$  in this interval of $I_i$. 
The QD current noise cumulant $\langle \langle J^2 \rangle
\rangle_{c}$ conditioned on the QPC current $I$ is obtained by 
$\langle \langle J^{2}\rangle
\rangle_{c}=t[\sum_{i=1}^{S}J_{i}^{2}/S-\langle \langle J\rangle
\rangle_{c}^{2}]$. 

The results of $\langle \langle J \rangle
\rangle_{c}$ and $\langle \langle J^{2}\rangle
\rangle_{c}$ obtained by the quantum trajectory method with    
120,000 realizations of $\langle n_2 \rangle_c(t)$ are shown in open
diamonds in 
Figs.~\ref{fig:SQD_cumulants}(c) and \ref{fig:SQD_cumulants}(d)
and in Figs.~\ref{fig:SQD_cumulants}(e) and
\ref{fig:SQD_cumulants}(f) 
for different sets of $(D,D')$ values.   
One notices that the results of the quantum trajectory method 
in Figs.~\ref{fig:SQD_cumulants}(e) and \ref{fig:SQD_cumulants}(f)
for large  
QPC tunneling rates or large $(D'- D)^2$
are in a better agreement with
those of other 
 methods  than  the results  
in Figs.~\ref{fig:SQD_cumulants}(c) and \ref{fig:SQD_cumulants}(d)
for small QPC tunneling rates or small $(D'- D)^2$.
This is because larger difference of QPC tunneling rates or larger
$(D'- D)^2$ represents a
better occupation number measurement of dot 2 and a better condition
to ignore the QPC shot noise,
which in turn give a    
more accurate number  $M$ 
of electrons that have tunneled through dot 2 into the right
lead of the QD system in time $t$ 
for a given trajectory 
 $\langle n_2 \rangle_c(t)$ 
and thus the corresponding average 
QD current $J={M}/{t}$.

\begin{figure}
\includegraphics[width=8.5cm]{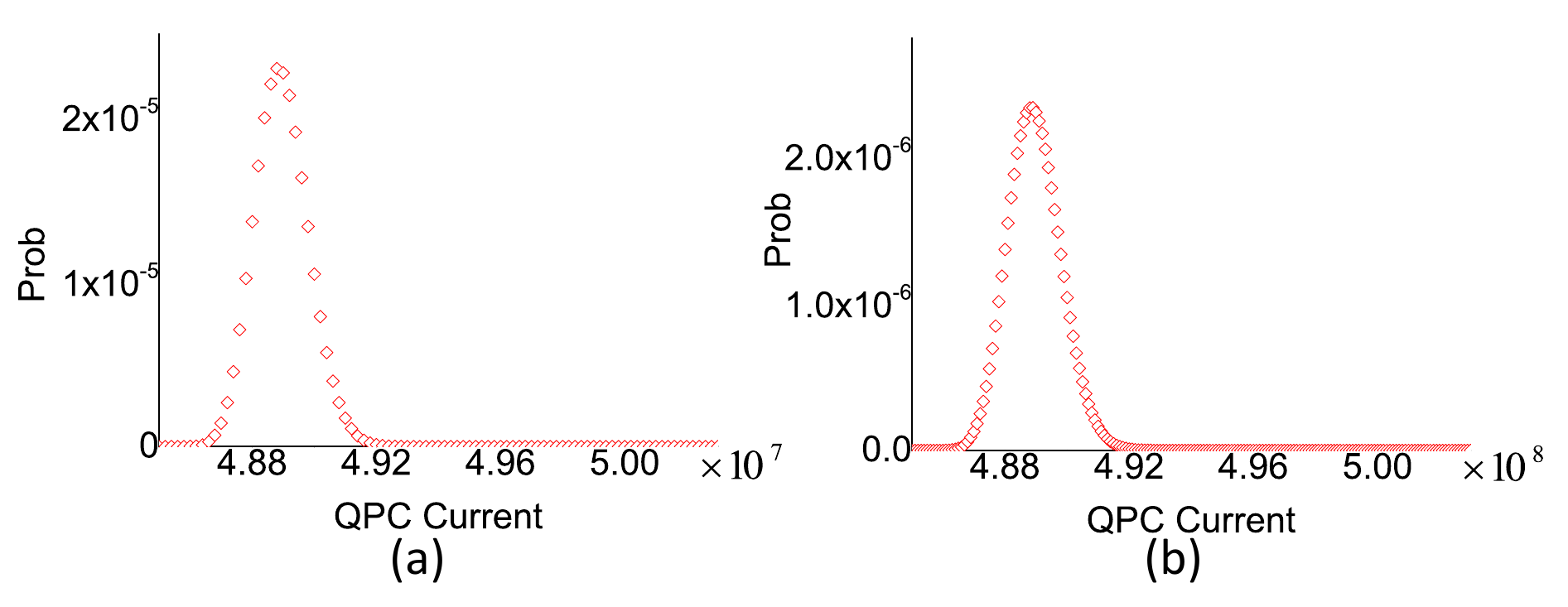}
\caption{Probability distributions of the QPC current 
(a) for the case presented in Figs.~\ref{fig:SQD_cumulants}(c) and ~\ref{fig:SQD_cumulants}(d),
and (b) for the case presented in Figs.~\ref{fig:SQD_cumulants}(e) and ~\ref{fig:SQD_cumulants}(f).
The whole range $[D,D']$ of the horizontal axis of the QPC current
shown in (a) and in (b)
is divided into 200 intervals.
}
\label{ProbI}
\end{figure}

One also notices that the open diamonds representing 
$\langle \langle J \rangle \rangle_{c}$ and 
$\langle \langle J^{2}\rangle \rangle_{c}$  
obtained by the quantum trajectory method  in 
Figs.~\ref{fig:SQD_cumulants}(c) and \ref{fig:SQD_cumulants}(d)
and in Figs.~\ref{fig:SQD_cumulants}(e) and
\ref{fig:SQD_cumulants}(f) 
show only
results in certain ranges 
of the QPC currents, which should correspond to the regimes where  
the probability distributions of the QPC current $P(I)$ are
not small. 
Indeed, the corresponding probability distributions $P(I)=\int P(I,J)dJ$ 
shown respectively in
Figs.~\ref{ProbI}(a) and Fig.~\ref{ProbI}(b) 
are highly
concentrated with appreciable values only in the same 
 small ranges of the QPC currents $I$
of their corresponding plots  
in Fig.~\ref{fig:SQD_cumulants}. 
Consequently, even
a large number of 120,000 quantum trajectories gives only data samples in
that small range of $I$. The first and last open
diamonds in $\langle\langle J^2 \rangle\rangle_c$ of
Figs.~\ref{fig:SQD_cumulants}(f) deviate more from the results    
obtained by other methods due to the fact that the numbers of
QD current data samples in the two 
corresponding QPC current intervals are not large enough to give
accurate statistics of conditional noise $\langle\langle J^2
\rangle\rangle_c$. We thus disregard the results of the QPC 
current intervals 
outside the regime bounded by the two intervals as less 
data samples are expected and observed. 

In short, the quantum trajectory method can, in principle, give the full
information about the transport properties provided a very
large number of trajectories are available.   
However, due to highly concentrated QPC current probability
distribution, a substantial amount of different realizations of
quantum trajectories that can perhaps already simulate
unconditional quantities well can still not sample the complete range of
the QPC currents for the conditional quantities. 
Thus an efficient method to calculate the conditional
counting statistics is demanding, and     
the method that is also capable of treating 
 more complicated nanostructure transport systems
(e.g., the DQD-QPC system)
described in Sec.~\ref{sec:EffNumMeth} provides
exactly 
such a method.

\begin{figure}
\includegraphics[width=8.5cm]{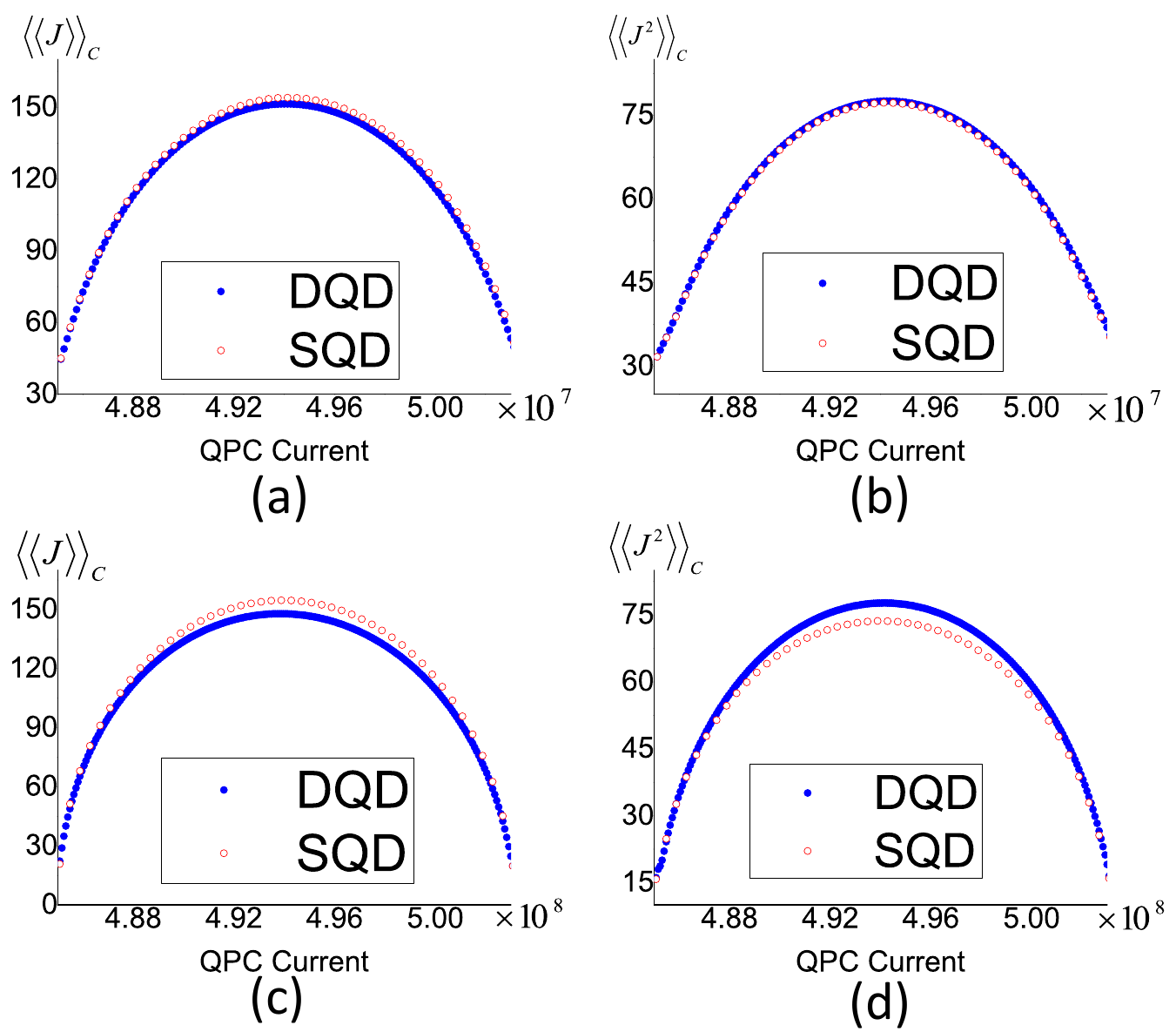}
\caption{Conditional DQD $\left\langle \left\langle J\right\rangle
  \right\rangle _{c}$ (left panel) and
$\left\langle \left\langle J^{2}\right\rangle \right\rangle _{c}$
(right panel) plotted as a function of the observed average QPC current 
$I$ with high interdot coupling $\Omega=15000$ Hz
for  
(a) and (b) $(D',D)=(5.03\times10^{7},4.85\times10^{7})$ Hz, and
(c) and (d) $(D',D)=(5.03\times10^{8},4.85\times10^{8})$ Hz.
The DQD cumulants plotted in solid blue dots are compared with their SQD
counterparts in open red dots with the same 
QD tunneling rates of $(\gamma_{L},\gamma_{R})=(160,586)$ Hz from the
left lead and to the 
right leads, respectively.}
\label{fig:DQD_large}
\end{figure}

\begin{figure}
\includegraphics[width=8.5cm]{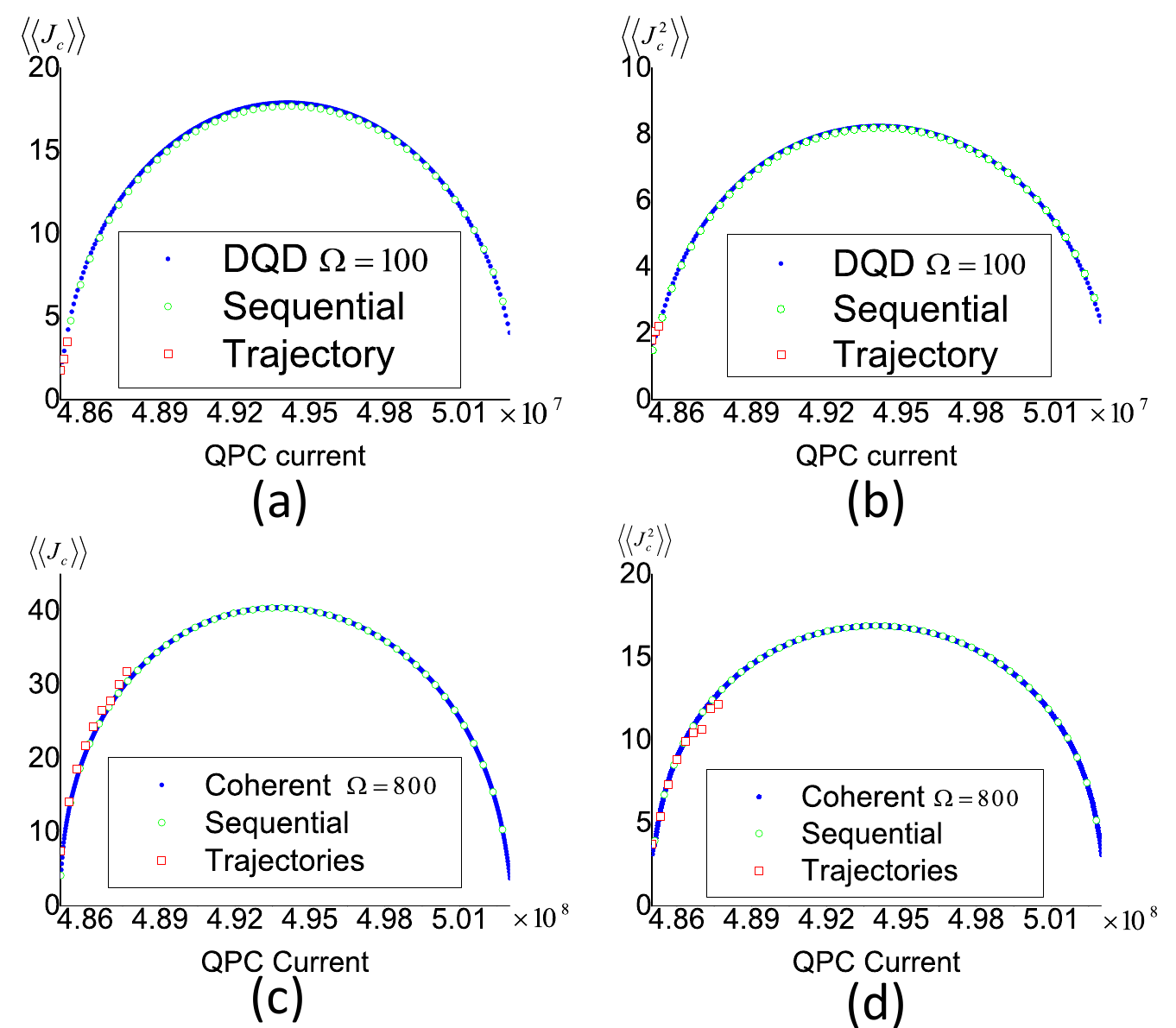}
\protect\caption{Conditional DQD $\left\langle \left\langle J\right\rangle
  \right\rangle _{c}$ (left panel) and
$\left\langle \left\langle J^{2}\right\rangle \right\rangle _{c}$
(right panel) plotted as a function of the observed average QPC current 
$I$ for different interdot couplings and QPC tunneling rates of
(a) and (b)
$(D',D,\Omega)=(5.03\times10^{\text{7}},4.85\times10^{7},100)$ Hz, and 
(c) and (d)
$(D',D,\Omega)=(5.03\times10^{\text{8}},4.85\times10^{8},800)$ Hz
obtained by the number-resolved master equations of coherent tunneling
(in blue solid dots)
and sequential tunneling (in open green dots) and by the method of
quantum trajectories (in open red squares). 
The tunneling rates of the DQD's 
 from the left lead and to the 
right leads are, respectively, $(\gamma_{L},\gamma_{R})=(160,586)$ Hz.
}
\label{fig:sequential}
\end{figure}
   
\noindent 
\begin{figure}
\includegraphics[width=8.5cm]{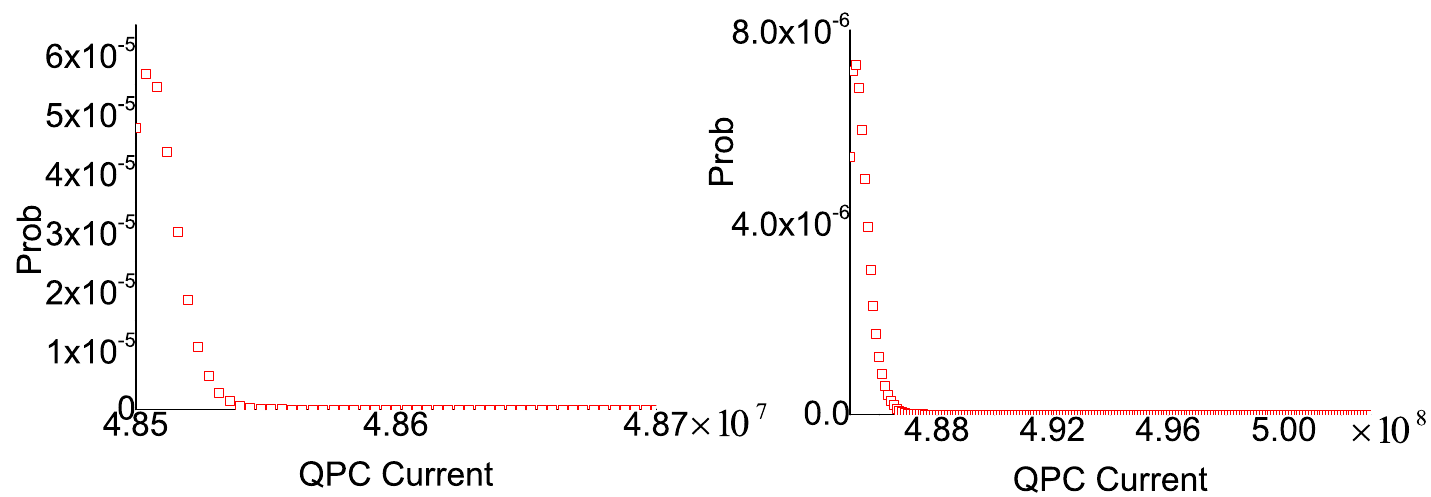}
\protect\caption{
Probability distributions of the QPC current
(a) for the case presented in Figs.~\ref{fig:sequential} (a) and (b),
and (b) for the case presented in Figs.~\ref{fig:sequential} (c) and
(d). The inset in (a) is its zoom-in plot for small QPC currents.
The whole range $[D,D']$ of the horizontal axis of the QPC current
shown in (a) and in (b)
is divided into 200 intervals.
}
\label{fig:Prob_sequential}
\end{figure}

\begin{figure}
\includegraphics[width=8.5cm]{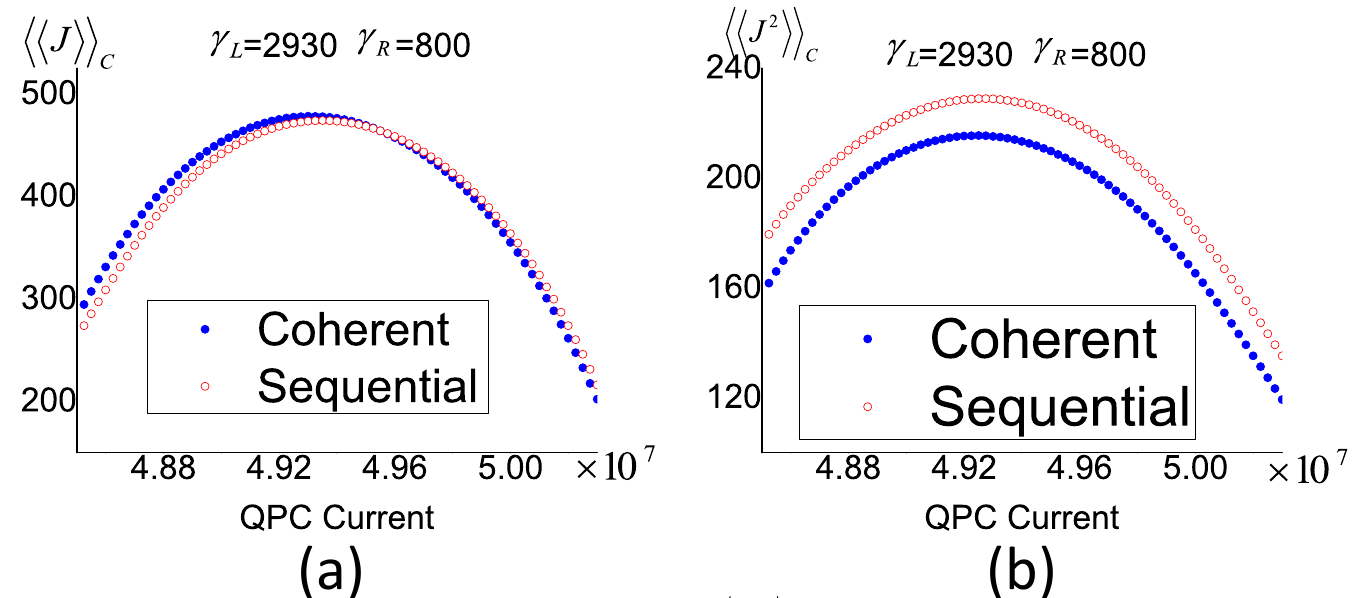}
\protect\caption{Conditional DQD $\left\langle \left\langle J\right\rangle
  \right\rangle _{c}$ (left panel) and
$\left\langle \left\langle J^{2}\right\rangle \right\rangle _{c}$
(right panel) plotted as a function of the observed average QPC current 
$I$ obtained by the number-resolved master equations of coherent tunneling
(in blue solid dots)
and sequential tunneling (in open red dots)
with interdot coupling $\Omega=3000$ Hz, QPC tunneling rates 
$(D',D)=(5.03\times10^{7},4.85\times10^{7})$ Hz, and
QD tunneling rates $(\gamma_{L},\gamma_{R})=(2930,800)$ Hz.
}
\label{fig:comparable}
\end{figure}

\subsection{DQD- QPC system}

Electron transport properties through a DQD system have been 
studied intensively 
\cite{Kouwenhoven02,Gurvitz96,Gurvitz97,Nazarov99,Schoell04,Schoell06,key-32,key-9,key-34,key-35},
Unconditional transport properties of a DQD system measured by a QPC 
have also been investigated
\cite{Gurvitz97,Schoell06,Gustavsson09,Marcus04,key-36,key-38,key-39}.
Here we concentrate on the conditional current cumulants through a DQD
system conditioned on the observed average QPC current,
which has not yet been explored extensively in the
literature.

\subsubsection{Conditional counting statistics}

As mentioned, it is numerically 
unstable to follow the same procedure \cite{Sukhorukov07} of 
taking (partial) derivatives to find conditional current cumulants
from Eqs.~(\ref{lambda_cI}) 
(\ref{lambda_cJ}), (\ref{eq:conditionalCCJ}) and
(\ref{eq:conditionalCCI}) 
for the DQD-QPC system. 
It is also numerically inefficient (using too much memory) 
to calculate the joint probability
distribution, Eq.~(\ref{eq:LT}), and then use the Bayesian
formalism to find the conditional quantities for the DQD-QPC system. 
 
Here, we use our numerically stable and efficient method to calculate 
$\left\langle \left\langle J\right\rangle \right\rangle _{c}$ and
$\left\langle \left\langle J^{2}\right\rangle \right\rangle _{c}$,
and discuss their dependence on $\Omega$ and $\Gamma_{d}$. 
The dephasing rates $\Gamma_{d}$ 
on the DQD's due to the QD-occupation-dependent QPC
tunneling rates of  
$(D',D)=(5.03\times10^{7},\:4.85\times10^{7})$ Hz and
$(D',D)=(5.03\times10^{\text{8}},\:4.85\times10^{8})$ Hz that will be
considered 
are estimated from Eq.~(\ref{eq:dephasing_rate}) to be about   
$\Gamma_{d}=8200$ Hz, and $\Gamma_{d}=82000$ Hz, respectively.  
We will discuss cases with 
 values of interdot coupling
$\Omega$ greater than, smaller than
and comparable to $\Gamma_d$.  

Considering the case of large  $\Omega=15000$ Hz,
we compare the conditional cumulants 
$\left\langle \left\langle J\right\rangle \right\rangle _{c}$ and
$\left\langle \left\langle J^{2}\right\rangle \right\rangle _{c}$
of the DQD's to those of SQD with the same values of $\gamma_{L}=160$ Hz
and $\gamma_{R}=586$ Hz as the DQD's.
In Figs.~\ref{fig:DQD_large}(a) and \ref{fig:DQD_large}(b) where 
$\Gamma_{d}=8200$ Hz is smaller than $\Omega=15000$ Hz, the agreement
in
$\left\langle \left\langle J\right\rangle \right\rangle _{c}$ and
$\left\langle \left\langle J^{2}\right\rangle \right\rangle _{c}$ 
between DQD's and SQD is rather good, indicating that
the coherently coupled DQD's can be approximately regarded as a SQD.
On the other hand, 
in Figs.~\ref{fig:DQD_large} (c) and~\ref{fig:DQD_large}(d) where $\Gamma_{d}=82000$ Hz is larger
than $\Omega=15000$ Hz, there are appreciable differences between 
DQD and SQD results 
due to the  effect of stronger dephasing (back action) caused by the QPC.
The larger 
$\Gamma_d$ 
tends to reduce the coherent
tunneling amplitude between the DQD's and hence reduce the current
passing through 
the DQD's. As a result, the value of the 
conditional current $\left\langle \left\langle J\right\rangle
\right\rangle _{c}$ of the DQD-QPC system 
is slightly smaller than that of the SQD-QPC system 
[see Fig.~\ref{fig:DQD_large} (c)]. 
In contrast,  the value of the 
conditional zero-frequency current noise $\left\langle \left\langle
    J\right\rangle\right\rangle _{c}$ of the DQD-QPC system is
slightly larger than that of regarding the DQD system as a strongly
coherently coupled SQD-QPC system [see
Fig.~\ref{fig:DQD_large} (d)].   This is consistent with
the unconditional noise property that quantum coherence suppresses noise 
 \cite{Schoell06,key-32,key-33}.

In the low coherent tunneling regime where $\Omega \ll \Gamma_{d}$,  
the QPC charge detector introduces substantial 
decoherence to the DQD's. Thus  
the dynamics of the electron transport of the coherent tunneling
 DQD's described by the $(6\times 6)$ matrix 
$\mathcal{L}(k,q)$ defined in Eq.~(\ref{eq:Lmatrix}) can be
in this case effectively described by a sequential tunneling $(4\times
4)$ matrix  
defined in Eq.~(\ref{eq:L_sequential}).
This is clearly shown in Fig.~\ref{fig:sequential} in which the
condition $\Omega \ll \Gamma_{d}$ holds and the results of $\left\langle \left\langle
    J\right\rangle \right\rangle _{c}$ and
$\left\langle \left\langle J^{2}\right\rangle \right\rangle _{c}$
calculated by our numerical method 
with the coherent tunneling matrix of Eq.~(\ref{eq:Lmatrix})
and with  the sequential tunneling
matrix of Eq.~(\ref{eq:L_sequential}) coincide.
Also shown in red open squares are the results obtained 
by the quantum trajectory
method, which are in good agreement with the results by our numerical
method. However, only the results conditioned on small values of 
$I_{QPC}$ are available due to small effective sequential tunneling rate
$\Gamma_{\Omega}$ of Eq.~(\ref{eq:Gamma_Omega}).
This then makes the second dot (the right dot) of
the DQD's preferring to be empty resulting in probability distribution 
of $P(I)$ concentrating in a small regime of the QPC currents $I$ as
shown in Figs.~\ref{fig:Prob_sequential}(a) and \ref{fig:Prob_sequential}(b).
This highlights the inability of the quantum trajectory method to
cover in practice the whole range of the QPC current $I$ for the cases
of extremely small $\Gamma_{\Omega}$.

However, when $\Omega$, $\Gamma_{d}$,  $\gamma_{L}$
and $\gamma_{R}$ are comparable,
using the classical master equation of the sequential
tunneling matrix  
of Eq.~(\ref{eq:L_sequential})
involving only the occupation probabilities cannot treat
this case of the DQD system. 
The unconditional steady-state currents
$\left\langle \left\langle J\right\rangle \right\rangle $ obtained by
the unconditional master equation of sequential tunneling 
and the unconditional quantum
master equation of coherent tunneling 
are the same independent of the DQD and QPC parameters
\cite{Schoell06,key-32,key-33}. 
But the steady-state unconditional zero-frequency noise  
$\left\langle \left\langle J^{2}\right\rangle \right\rangle $
in the parameter regime in which $\Omega$, $\Gamma_{d}$,  $\gamma_{L}$
and $\gamma_{R}$ are all comparable \cite{Schoell06,key-32,key-33},
shows considerable difference between the coherent and the sequential
tunneling cases. 
Choosing  comparable parameters of interdot coupling $\Omega=3000$ Hz,
dephasing rate by the QPC $\Gamma_{d}=8200$ Hz for the QPC currents 
$(D',D)=(5.03\times10^{\text{7}},4.85\times10^{7})$ Hz, and the DQD
tunneling rates from the left lead and to the right leads 
$(\gamma_{L},\gamma_{R})=(2930,800)$ Hz, respectively, we show in
Fig.~\ref{fig:comparable} the conditional
steady-state DQD current and zero-frequency 
noise for  $\triangle\epsilon=0$.
One can see that the steady-state conditional currents 
$\left\langle \left\langle
    J\right\rangle \right\rangle _{c}$ obtained by the coherent-tunneling and
sequential-tunneling master equations show still some observable
difference, in contrast to no difference in their 
unconditional counterparts. 
The conditional steady-state noise 
$\left\langle \left\langle J^{2}\right\rangle \right\rangle_c $ obtained
by the coherent-tunneling master equation with the matrix of
Eq.~(\ref{eq:Lmatrix})
is considerably 
smaller than that obtained by the
sequential-tunneling master equation with the matrix  
of Eq.~(\ref{eq:L_sequential})
, i.e., quantum coherence
suppresses noise as in the unconditional case \cite{Schoell06,key-32,key-33}.

The unconditional current moment 
can be expressed as $\left\langle J^r\right\rangle
=\intop_{0}^{\infty}dI\, P(I)\left\langle J^{r}\right\rangle _{c}$ and thus
 can be calculated from the conditional counting statistics.
We thus use the
conditional counting statistics obtained by our numerical method to 
compute unconditional
current cumulants.  
The results are consistent with the unconditional current cumulants
obtained by the method of Ref.~\onlinecite{Schoell06}. Thus 
conditional counting
statistics can provide more detailed information about and physical insight 
into the quantum transport properties of the system than
its  unconditional counterpart.

\section{Conclusion}
\label{sec:Conclusion}
We have applied the maximum eigenvalue method, the quantum trajectory
method and a stable and efficient method we develop to calculate the
conditional counting statistics of QD systems measured by a QPC
detector. The method we develop is capable of calculating
the conditional counting statistics for a more complex system than the maximum eigenvalue method
and for a wider range of parameters than the quantum trajectory method.
We have investigated the effect
of QPC shot noise on the conditional cumulants of the
QD systems. 
For the considered case of high QPC tunneling rates for which the
QPC shot noise as compared to the noise contribution
of the random telegraph signal in the QPC current trace
is small and can be neglected, our results are in 
excellent agreement with
those obtained by the method of Ref.~\onlinecite{Sukhorukov07}.
However, for the cases of low QPC tunneling rates for which the
QPC shot noise cannot be neglected, significant difference 
between
$\left\langle \left\langle J\right\rangle \right\rangle _{c}$ and
$\left\langle \left\langle J^{2}\right\rangle \right\rangle _{c}$
obtained by our method and those obtained 
by the analytical solutions of Ref.~\onlinecite{Sukhorukov07} can be
observed.  
We have also shown that 
for strong interdot coupling of $\Omega\gg\Gamma_{d}$, conditional DQD cumulants
 $\left\langle \left\langle J\right\rangle \right\rangle _{c}$ and
$\left\langle \left\langle J^{2}\right\rangle \right\rangle _{c}$
recover those of a SQD case (i.e., the DQD's acct as a SQD).
For small interdot coupling ($\Omega\ll\Gamma_{d}$),
the results of  $\left\langle \left\langle J\right\rangle \right\rangle _{c}$ and
$\left\langle \left\langle J^{2}\right\rangle \right\rangle _{c}$
calculated
by our numerical method with the coherent-tunneling
matrix and with the sequential-tunneling matrix coincide, while they
show considerable difference when $\Omega$, $\Gamma_{d}$,  $\gamma_{L}$,
and $\gamma_{R}$ are all comparable.
The conditional current cumulants
that are significantly different from their unconditional counterparts
can provide additional information and insight into the electron  
transport properties of mesoscopic nanostructure systems.

\begin{acknowledgments}
H.S.G. acknowledges support from the the Ministry of Science and Technology
of Taiwan under Grants No.~MOST 103-2112-M-002-003-MY3 and No.~MOST
106-2112-M-002-013-MY3, from the National
Taiwan University under Grant No.~NTU-CCP-106R891703
and from the thematic group program of the National Center for Theoretical
Sciences, Taiwan.
We also acknowledge useful discussion with Dr. Chien-Hung Lin
in the early 
stage of this work.
\end{acknowledgments}

\appendix
\section{Semiempirical method to count the electron tunneling events}
\label{App:counting_method}
The detailed procedure of the semiempirical method
used to count the number of tunneling electrons in each of 
the quantum trajectories 
is described as follows.   
First, divide the entire range of values from 0 to 1 of $\left\langle
  n_2\right\rangle _{c}(t)$ 
into several small intervals and count the numbers of $\left\langle
  n_2\right\rangle _{c}(t)$ values that fall into each
interval in a quantum trajectory (see
Fig.~\ref{fig:histogram}). Second, select 
the interval whe re the value of histogram is minimum and let the middle value of
$\left\langle n_2\right\rangle _{c}(t)$ in this minimum interval as
the reference 
value or line [as indicated in Fig.~\ref{fig:histogram} (b)]
 to determine whether the QD is occupied or empty.  Third,
calculate the average value of $\left\langle n_2\right\rangle_c $ over
the region on the left (right) side of the reference line and set it
as the lower (upper) threshold
whose value is usually close to 0 (1). 
Suppose the QD is initially being empty, 
the value of  $\left\langle n_2\right\rangle _{c}(t)$ is small.
Then the QD is considered being occupied only until  
 $\left\langle n_2\right\rangle _{c}(t)$  is above the upper threshold;
the QD is considered being empty again only until 
$\left\langle n_2\right\rangle _{c}(t)$ is below the lower threshold.
Thus when $\left\langle n_2\right\rangle _{c}(t)$ in a quantum
trajectory realization 
reaches the above occupied situation and then the immediate empty
 situation sequentially and successively, an
electron tunneling event 
from the QD to  the right lead is registered.  By this counting
method, we can obtain 
the number $M$ of electrons that have tunneled through the QD system.
With this counting method, one can proceed to calculate the
conditional current cumulants as described in Sec.\ref{sec:CCSbyQT}.

\end{document}